\documentclass[11pt]{article}
\usepackage[final]{graphics}
\usepackage{amsfonts,amsbsy}

\begin{document}

\thispagestyle{empty}
\title {\bf The Color Glass Condensate:\\
{\small A summary of key ideas and recent developments}   }

\author{Raju Venugopalan$^{1,2}$}
\maketitle
\begin{center}
\begin{enumerate}
\item Physics Department\\
  Brookhaven National Laboratory\\
  Upton, NY 11973, USA
\item Physics Department\\
Bielefeld University\\
D-33615, Bielefeld, Germany
\end{enumerate}
\end{center}

\begin{abstract}
We summarize  the theory and phenomenology of the Color Glass Condensate reviewed previously by E. Iancu and the author in hep-ph/0303204. In addition, we discuss some of the subsequent developments in the past year both in theory and in phenomenological applications. 

\end{abstract}

\section{Introduction}

In the spring of this year, I delivered three 90 minute lectures on the Color Glass Condensate (CGC) at the joint Hadron 2004/RANP meeting of the 
Brazilian high energy and nuclear physics communities held at Angra dos Reis in the state of Rio De Janeiro, Brazil~\footnote{Subsequently, a version 
of these lectures was presented at HUGS 2004 at Jefferson Lab, and at the University of 
Bielefeld. Part of the material discussed here was also presented at the CTEQ summer school in Madison, Wisconsin and at the Erice conference {\it QCD at cosmic energies}.}. A sizeable 
fraction of the material covered there was discussed previously by E. Iancu and myself in a review last year~\cite{IV}. I will therefore only 
summarize those topics and refer the reader to Ref.~\cite{IV} for details. This summary may be useful to the reader who might be interested in a quick review of the subject. Subsequent to the review in Ref.~\cite{IV}, 
there have been several important developments in both the theory and phenomenology of the CGC. (The latter in particular has been 
greatly stimulated by the remarkable data on dA scattering taken by the RHIC experiments~\cite{RHIC_dA}.) Some of these 
developments were discussed at Angra by myself as well as by Iancu~\cite{Iancu1} and McLerran~\cite{McLerran1}. We 
shall discuss these developments wherever appropriate. Hopefully these lectures will then present an up to date perspective 
on the CGC for interested readers. 
A few other reviews which overlap with the material covered in these lectures are listed in Ref.~\cite{reviews}.

\subsection{Summary of Three Lectures on the CGC}

The outline of these lectures is as follows. In Lecture I, we  begin with a general introduction to outstanding problems in high 
energy QCD. We introduce the paradigm of Deeply Inelastic Scattering (DIS) and in this context discuss the evolution 
equations of perturbative QCD. A likely consequence of QCD evolution at small x is saturation of the 
wavefunction~\cite{GLR,MuellerQiu} in the 
limit~\footnote{Here $Q^2= -q^2 > 0$, where $q^2$ is the momentum transfer squared in deeply inelastic scattering, and 
$s=(P+k)^2$ is the center of mass energy squared in the collision of an electron with four momentum $k$ and a hadron 
with four momentum $P$.  The Bjorken variable $x$ is defined as $x = Q^2/2P\cdot q$, and one has $x\,y\approx Q^2/s$, 
where $y= P\cdot q/P\cdot k$ is called the inelasticity. For simplicity, we will assume throughout that $y=1$ and hence 
$x\approx Q^2/s$.} of $Q^2$ = fixed and $s\rightarrow \infty$. This leads to the formation of a Color Glass Condensate~\cite{MV,JIMWLK} for $Q^2 < Q_s^2(x)$, 
where $Q_s$ is the saturation scale. We discuss a simple model 
of DIS, the Golec-Biernat--Wusthoff model~\cite{GBW1}, which provides a saturation inspired phenomenology of DIS. 

In Lecture II, we begin with a discussion of key features of Light Cone quantized QCD. In particular, the parton model picture of QCD is 
manifest in Light Cone quantization. It provides a consistent framework to discuss the multi-gluon components of the infinite momentum frame 
wavefunction that dominate high energy scattering in QCD. This discussion provides the background and the motivation to construct a 
classical effective field theory that describes the ground state properties of hadrons at high energies~\cite{MV}. We next discuss the small x 
renormalization group equations (often called JIMWLK equations) which consistently include quantum corrections to the classical effective theory as one goes to higher and higher energies~\cite{JIMWLK}. The full structure of solutions to the JIMWLK equations is extremely complicated and only 
mean field analytical solutions are known~\cite{IIM}. A simplified equation which captures some of the non-linear structure of the 
full RG evolution is the Balitsky-Kovchegov (BK) equation~\cite{BK}. The BK equation has not been solved 
analytically to date though numerical solutions of 
these equations exist~\cite{Braun,ArmestoBraun,GolecStastoMotyka}. These are discussed briefly. The BK equation can be solved in a 
"diffusion" approximation~\footnote{It was later argued by these authors (in the second and third papers of Ref.~\cite{MunierPeschanski}) that the analogy with travelling waves goes 
beyond this approximation and is generic of saturation models in QCD. This is because in the latter, like the former, 
is characterized by unstable, rapid linear growth that is damped by non-linear effects which lead to saturation.}in analogy with the theory of travelling wave fronts in statistical physics~\cite{MunierPeschanski}. The analogy is a powerful one and very recently it has led to interesting developments in the treatment of fluctuations at high energies~\cite{IMM,IT}.

Lecture III deals with the application of the CGC picture to describe pp, p/D-A collisions and A-A collisions. We begin by briefly describing theoretical attempts in the CGC/saturation picture to fit the HERA DIS data. We next briefly discuss the validity of $k_\perp$ factorization in 
hadronic scattering at high energies. This is of phenomenological interest because some saturation models employ $k_\perp$ factorization 
and others do not. The recent RHIC data on Deuteron-Gold collisions has remarkable features especially at forward rapidities. We discuss the CGC interpretation of 
this data and argue that the unusual features of the data at forward rapidities have a natural explanation in the CGC framework.  Finally, we discuss CGC based approaches to the phenomenology of heavy ion collisions. While there is an emerging consensus~\cite{MiklosLarry} that final state interactions are essential to understand the RHIC data,  the initial state (as described by the CGC) helps constrain the properties 
of the bulk matter formed in heavy ion collisions. An outstanding theoretical problem is whether thermalization is achieved starting 
from the CGC. There has been much recent excitement about this possibility and we shall briefly summarize that discussion.

\section{Lecture I}

Perturbative QCD is very successful but it describes only a very small part of the total cross-section. Recall 
that 
\begin{equation}
\sigma_{\rm Rutherford} \propto {1\over Q^2} \, ,
\label{eq:1}
\end{equation}
which is negligibly small in the limit of $Q^2\rightarrow \infty$. The bulk of the semi-hard and soft cross-section remains to be understood. 
Lattice QCD is a first principles approach but it has been successfully applied so far only to static quantities. Its applicability is least at 
high energies where higher moments of leading twist operators become as important as lower moments, and moreover, where higher twist operators 
begin to contribute significantly to physical processes. The fact that the two most rigorous techniques, perturbation theory and the lattice, are 
inadequate is especially unfortunate because it appears that there may exist simple structures controlling the high energy behavior of cross-sections. 
These regularities have been known for a long time and understanding these in the language of the fundamental theory remains an outstanding 
scientific puzzle. One such example is the parametrization of total cross-sections as 
\begin{equation}
\sigma(s) = As^{0.0808} + Bs^{-0.4525} \, ,
\label{eq:2}
\end{equation}
advocated by Donnachie and Landshoff~\cite{DL}.  The first term corresponding to rising 
cross-sections  with energy is believed to occur due to the exchange of 
a particle with vacuum quantum numbers (the Pomeron), while the second term (corresponding 
to falling cross-sections) is due to Reggeon exchange. Such simple fits work extremely well for a large 
number of observables. It must be noted that alternative parametric forms are argued to 
work just as well. Assuming that the Donnachie-Landshoff fits, by the logic of Occam's razor, correspond 
to reality, what are Pomerons and Reggeons and how can we construct them from the underlying 
theory? The most sophisticated attempts to understand the Pomeron involves the exchange of 
two "reggeized gluons" -the Pomeron thereby constructed is called the BFKL Pomeron~\cite{BFKL}. 
The BFKL Pomeron is constructed in perturbation theory (very sophisticated perturbation theory) 
but it has a few problems. More sophisticated work can cure some of these problems but the 
fact remains that the BFKL Pomeron does not describe total cross-sections and indeed is hard pressed to explain perturbative results as well. 

\begin{figure}[htbp]
\begin{center}
\resizebox*{!}{4.0cm}{\includegraphics{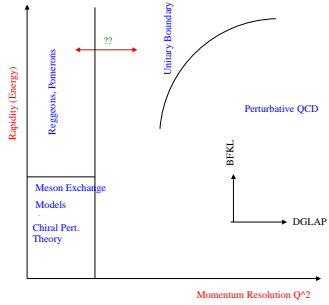}}
\end{center}
\caption{\label{fig:one}A road map of the strong interactions. }
\end{figure}

A ``road map" of the strong interactions shown in Fig.~\ref{fig:one} is useful at this stage. On the y-axis, we plot the Rapidity (or 
the Energy) and on the x-axis the momentum resolution ($Q^2$). At low $Q^2$ and low energies, 
Chiral Perturbation Theory works very well. Meson exchange modes do a reasonable job at 
somewhat higher energies (but still small values of $Q^2$). At high energies, and small momentum 
transfers, Pomeron phenomenology is successful if not, as discussed, understood. On the other 
extreme of the plot, the perturbative QCD evolution equations work very well. In perturbative QCD, 
one obtains large logarithms in $\alpha_s\ln(Q^2/Q_0^2)\ln(x_0/x)$, where $x_0$ and $Q_0^2$ 
correspond to the initial values in the evolutions. The DGLAP equations~\cite{DGLAP} compute the leading logarithms in $\alpha_s\ln(Q^2/Q_0^2)$, while the logarithms in $\alpha_s\ln(x_0/x)$ are 
sub-leading. The situation is reversed for the BFKL equation, where the leading logarithms are those in $\alpha_s\ln(x_0/x)$. Both approaches lead to linear evolution equations, in $\ln(Q^2/Q_0^2)$ 
and $\ln(x_0/x)$ respectively, and both lead as well to rapidly rising cross-sections at small x.  Very rapidly rising cross-sections violate unitarity and there must therefore exist a mechanism in the theory which restores 
unitarity at small x. Since the unitary boundary may be reached even for $Q^2>>\Lambda_{\rm QCD}^2$ at sufficiently small x, many of us believe that the dynamics that leads to unitarization can be 
understood in weak coupling~\cite{Mueller1}. It is interesting to speculate whether the physics of unitarization via weak coupling may also provide insight into the successful phenomenology of high energy scattering in 
the language of Pomerons and Reggeons.  Historically, this physics was believed to be entirely non-perturbative in nature.

\begin{figure}[htbp]
\begin{center}
\resizebox*{!}{4.0cm}{\includegraphics{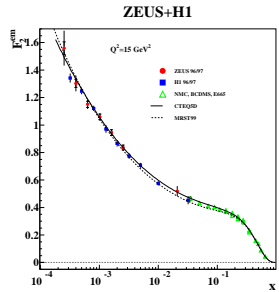}}
\end{center}
\caption{\label{fig:two}H1 and ZEUS QCD fits to $F_2$ data for fixed $Q^2$ as a function of $x$. }
\end{figure}

\begin{figure}[htbp]
\begin{center}
\resizebox*{!}{4.0cm}{\includegraphics{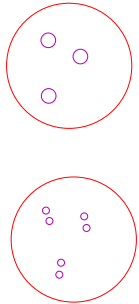}}
\end{center}
\caption{\label{fig:three} Phase space density from DGLAP evolution.}
\end{figure}

DIS experiments at HERA ushered in a new era in high energy QCD when they showed conclusively that structure functions rise very rapidly at small x. It came as a surprise (even though it shouldn't have), since 
predictions for this behavior existed right from the very early days of QCD~\cite{Gross}. It has since been shown that DGLAP based QCD fits are very successful in fitting the HERA data~\cite{H1ZEUSQCD}. 
These are shown in Fig.~\ref{fig:two}. In DGLAP evolution, the cross-section grows as one increases $Q^2$. However, even though the multiplicity apparently grows with increasing $Q^2$, the phase space density actually decreases. 
This is because even though one sees many more partons, they are smaller in size. A cartoon demonstrating this is shown in Fig.~\ref{fig:three}. We would like to ask what would happen if one were to fix $Q^2$ and decrease x. In DGLAP, in part, this depends on the initial $x$ distribution, which has to be a fairly steep power law. A natural mechanism is given by BFKL evolution, which gives rise to the rapid growth in x similar to those assumed in DGLAP initial conditions. The phase space density in this case grows very rapidly. 
Thus the great value of the BFKL equation is that it provides a mechanism to rapidly generate large numbers of partons, which in turn give rise to large cross-sections~\footnote{Our 
perspective of BFKL here is in the S-channel dipole scattering picture of Mueller~\cite{Mueller2}}. 

\begin{figure}[htbp]
\begin{center}
\resizebox*{!}{4.0cm}{\includegraphics{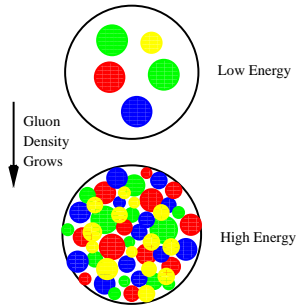}}
\end{center}
\caption{\label{fig:four} Phase space density from BFKL evolution.}
\end{figure}

A cartoon demonstrating this growth in the phase space density is shown in Fig.~\ref{fig:four}. When the phase space density becomes large, of order $1/\alpha_S$, partons fill up the entire phase space of the 
hadron saturating the phase space density and leading to the phenomenon known by this name~\cite{GLR}. 
Saturation is driven by higher twist effects corresponding to the screening and recombination of partons in 
the high density environment. These contributions are non-linear in the parton density and their  net effect is to slow down the rise in the parton density~\cite{MuellerQiu}.  These effects are not contained in BFKL evolution, which describes 
only linear evolution in the (unintegrated) parton densities.

The competition between the 
perturbative Bremsstrahlung and the many body, non-linear screening and recombination effects is characterized by a scale $Q_s(x)$ in the $x-Q^2$ plane. When $Q < Q_s(x)$ one expects the latter 
to dominate, while for $Q >> Q_s$, the linear evolution characteristic of QCD Bremsstrahlung is apparent. Saturation therefore provides a natural mechanism to unitarize cross-sections. (The converse is 
not necessarily true: a unitarized cross-section is not necessarily a saturated cross-section.)

The Color Glass Condensate is an effective field theory which describes the phenomenon of saturation in QCD~\cite{MV,JIMWLK}. 
The large parton density provides a large scale, the saturation scale $Q_s^2 \propto \alpha_s N_c 
dN/dy/\pi R^2$, which ensures that the physics can be described in weak coupling. The high occupation number, $n\sim 1/\alpha_S$, ensures that the small x modes can be treated as classical fields. These 
are coupled to static sources at larger values of x. The evolution of the source density with x is described by renormalization group equations. The strong fields in the CGC description ensure saturation already 
at the classical level of the theory. This is preserved by quantum evolution, which introduces additional non-trivial features. We will discuss the CGC further in Lecture II. 

At what values of x does saturation set in? This is a difficult question to answer from first principles~\footnote{It is no easier to answer this 
question than 
to state at what values of $Q^2$ perturbative QCD is applicable. The answer depends on what process one is calculating.}  since one does not have quantitative control over where the higher twist effects that contribute to 
saturation become important. The data presented in Fig.~\ref{fig:two} don't appear to show any saturation effects. However, there are indications from the behavior of the gluon distribution (at small x and moderately small $Q^2$) 
and more importantly, from the longitudinal structure function $F_L$ extracted from the data that higher twist effects may be playing a role already at HERA. It has been argued by Bartels, Golec-Biernat and Peters~\cite{BGP} that higher twist contributions that may be significant in $F_L$ and $F_T$ separately, cancel in $F_2$ (=$F_L+F_T$). Unfortunately, the 
$F_L$ data from HERA has large systematic errors at small x.

It is well known that the virtual photon-proton cross-section at small x can be written as~\cite{BjKogutSoper,Mueller90,NZ} 
\begin{equation}
\sigma_{T,L}^{\gamma^* p}=\int d^2 r_\perp \int dz |\psi_{T,L}(r_\perp,z,Q^2)|^2 \sigma_{q{\bar q}N}(r_\perp,x) \, ,
\label{eq:3} 
\end{equation}
where $|\psi_{T,L}|^2$ is the probability for a longitudinally or transversely polarized virtual photon to split into a quark with momentum fraction $z$  and 
an anti-quark with momentum fraction $1-z$ of the longitudinal momentum of the virtual photon. This probability is known exactly and it is convoluted with the cross-section for the $q\bar q$-pair to scatter off 
the proton. 
A simple model of small x DIS was introduced by 
Golec-Biernat and Wusthoff~\cite{GBW1}. In this model, the $q{\bar q} N$ cross-section (often called the dipole cross-section) was parametrized to be
\begin{equation}
\sigma_{q\bar q p}(r_\perp,x) = \sigma_0\,\left[1-\exp\left(-{r_\perp^2 Q_s^2(x)\over 4}\right)\right] \, ,
\label{eq:4}
\end{equation}
where 
\begin{equation}
Q_s^2(x) = Q_0^2 \left({x_0\over x}\right)^\lambda \, .
\label{eq:5}
\end{equation}
Golec-Biernat and Wusthoff  found a fit to all the HERA data for $x\leq 10^{-2}$ and $Q^2\leq 20$ GeV$^2$ with the parameters $Q_0=1$ GeV, $\sigma_0=23$ mbarns, $x_0=3\cdot 10^{-4}$ and $\lambda=0.3$. Interestingly, 
this model also provided a good fit to the diffractive data~\cite{GBW2}, in particular the surprising energy dependence of the diffractive cross-section. The form of the hadronic cross-section adopted 
by Golec-Biernat and Wusthoff arises naturally in the CGC picture~\cite{MV99}. The latter, as we shall discuss further in Lecture II, also has the right large $k_\perp$ behavior, the absence of which is 
one of the shortcomings of the Golec-Biernat-Wusthoff model. An improved version of the Golec-Biernat--Wusthoff model 
was developed by Bartels, Golec-Biernat and Kowalski~\cite{BGBK}, who matched the saturation model to DGLAP evolution at 
higher values of $Q^2$, while maintaining its key features. A further improvement of the model was the inclusion of the 
impact parameter dependence of the saturation scale by Kowalski and Teaney~\cite{KT}. An alternative approach 
was followed  by Mueller, Munier and Stasto~\cite{MMS}, who inferred the S-matrix (and therefore the saturation 
scale in the Golec-Biernat--Wusthoff parametrization) from the $t$-dependence of exclusive $\rho$-meson production. 
The value of the saturation scale extracted at small $x$ in all the approaches lies in the ball park of $Q_s^2\sim 1$-$1.5$ 
GeV$^2$. A similar analysis of $J/\psi$ production was performed by Guzey et al.~\cite{GRSZ}. For other saturation approaches, see 
Ref.~\cite{FSLM}. For a critical discussion of different models, see Ref.~\cite{RS}. These authors conclude that,
for the most central impact parameters, the parton densities are large enough at HERA to generate semi-hard 
scales which significantly impact the dynamics. An extensive recent theoretical 
review of vector meson production of HERA can be found in Ref.~\cite{INS}.

\begin{figure}[htbp]
\begin{center}
\resizebox*{!}{4.0cm}{\includegraphics{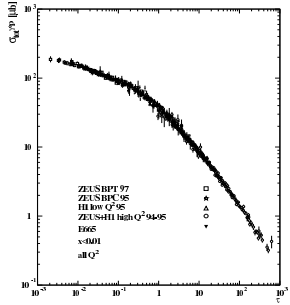}}
\end{center}
\caption{\label{fig:five} Geometrical scaling of the virtual photon-proton cross-section with $\tau = Q^2/Q_s^2$.}
\end{figure}

The Golec-Biernat model also inspired Golec-Biernat,   
Kwiecinski and Stasto~\cite{GKS} to plot the HERA data for the virtual photon-proton cross-section in terms of 
the dimensionless scaling variable $\tau= Q^2/Q_s^2$. Their result, as shown in Fig.~\ref{fig:five}, demonstrated that the data for $x<0.01$ and $0.045 < Q^2 < 450$ GeV$^2$ scaled very nicely as a 
function of $\tau$. Why these data scale up to these high $Q^2$ at small $x$ will be discussed in Lecture III. 

A key ingredient in the Golec-Biernat--Wusthoff model is the dipole cross-section $\sigma_{q{\bar q}p}$ in Eq.~\ref{eq:3}, 
for which they introduced the simple model in Eq.~\ref{eq:4}. We will see in the following that the dipole cross-section is 
ubiquitous and is a common ingredient in both DIS and hadronic scattering at high energies. 

\section{Lecture II}

In this lecture, we shall discuss an effective field theory for high energy scattering-the Color Glass Condensate. A key ingredient in our discussion will be the n-gluon component of the hadron wave function 
(where n is large!). In order to construct this wave function, we need to quantize the theory on the light cone. The initial value quantum problem is formulated on an equal {\it light cone time} surface 
$x^+ = (t+z)/\sqrt{2}=0$. The great advantage of this approach is that the vacuum simplifies greatly~\footnote{As the reader might suspect, there is no free lunch and 
the complications of the usual vacuum are transferred elsewhere. However for the purposes of perturbation theory, perhaps to our eventual peril, these may be neglected.} .
 For instance, the light cone quantized boost operator commutes with the 
light cone QCD Hamiltonian. This property is not satisfied by the usual equal time boost operator, which leads to particle creation under boosts. Thus if we wish to 
construct boost invariant wave functions, we must need do so on the light front. 

Light cone quantized quantum field theories have remarkable properties~\cite{LCreviews}. It was first noticed by Weinberg~\cite{Weinberg} that the Feynman rules of scalar field theories simplified greatly 
in the limit where the light cone momentum $P^+\rightarrow\infty$.  Subsequently, Susskind~\cite{Susskind} showed that there exists an exact isomorphism between the Poincare group of a quantum field 
theory on the light cone and the Galilean subgroup of two dimensional quantum mechanics. Thus for instance, as in quantum mechanics, the QCD light cone Hamiltonian ($P_{\rm QCD}^-$,  
the generator of translations in $x^+$), for instance, can be written 
as $P_{\rm QCD}^- = P_{\rm QCD}^{-,0} + V_{\rm QCD}$. One can construct a complete set of Fock states that are eigenstates of the free Hamiltonian. Now since the light cone vacuum is an 
eigenstate of both $P_{\rm QCD}^-$ and $P_{\rm QCD}^{-,0}$, one can write the physical eigenstate in terms of a complete basis of ``parton" eigenstates of  bare quanta. This simple observation (coupled with the physics of time dilation) is 
the basis of the parton model~\cite{Feynman} in quantum field theory. An elegant paper by Bjorken, Kogut and Soper~\cite{BjKogutSoper} which demonstrates these features in Quantum Electrodynamics on the light cone is 
recommended reading for anyone interested in understanding high energy scattering in a light cone quantized gauge theory.

In this light cone framework, the wavefunction of a high energy hadron, can therefore be expressed as 
\begin{equation}
|h> = |qqq> + |qqqg> + \cdots + |qqqg\cdots q\bar q ggg> \,,
\label{eq:6}
\end{equation}

\begin{figure}[htbp]
\begin{center}
\resizebox*{!}{4.0cm}{\includegraphics{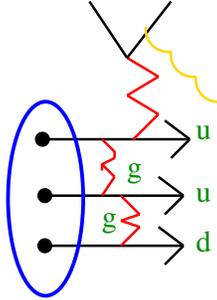}}
\end{center}
\caption{\label{fig:six} Wee partons in the infinite momentum frame.}
\end{figure}

Each wee parton in a configuration containing a large number of partons carries a small fraction  
$x= k^+/P^+$ of the total momentum $P^+$ of the hadron. These small x configurations (shown in 
Fig.~\ref{fig:six}) forming a fuzzy, strongly correlated parton cloud extending beyond the Lorentz contracted valence partons, 
can only be probed at high energies, and are the relevant configurations for multi-particle production. 
We cannot at present write down the wave function for these configurations in a hadron~\footnote{However, this is 
precisely what Al Mueller has done for the specialized case of the n-gluon component of a large onium pair in the large $N_c$ limit~\cite{Mueller2}.}. One can instead write down a path integral for the ground state of a hadron, 
which picks out the contributions from multi-parton states. 

\begin{figure}[htbp]
\begin{center}
\resizebox*{!}{4.5cm}{\includegraphics{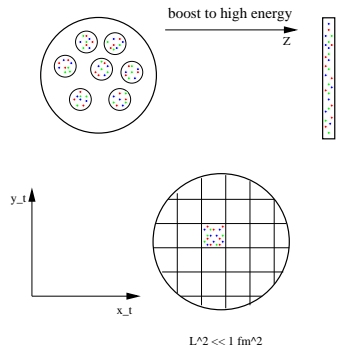}}
\end{center}
\caption{\label{fig:seven} A large nucleus boosted to high energies. A colored probe with transverse resolution $x_\perp\sim {1\over k_\perp}<< {1\over \Lambda_{\rm QCD}}$ sees a large number of color charges.}
\end{figure}

A classical effective theory for the small x modes of a large nucleus was constructed by Larry McLerran 
and myself~\cite{MV}. This theory is a coarse grained field theory constructed in the infinite momentum frame (IMF) $P^+\rightarrow \infty$ 
and in the light cone gauge $A^+=0$. In the IMF frame, only one component $J^+$ of the valence parton current is important-the others are suppressed 
by $1/P^+$. Also, as shown in the cartoon in Fig.~\ref{fig:seven}, due to Lorentz contraction, a wee parton 
with a transverse resolution $x_\perp\sim 1/k_\perp<<1/\Lambda_{\rm QCD}$ will see a large number of color charges from nucleons, all squished together in the transverse plane. 

\begin{figure}[htbp]
\begin{center}
\resizebox*{!}{5.5cm}{\includegraphics{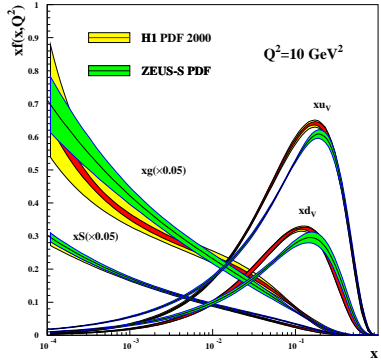}}
\end{center}
\caption{\label{fig:eight} The valence, glue and sea distributions in a proton. Note that the glue distribution is scaled 
down by a factor of 20!}
\end{figure}

Another important ingredient in our construction is the observation that the valence and glue (+ accompanying 
sea) modes are cleanly separated in x. This is seen in Fig.~\ref{fig:eight}. A remarkable feature of this plot is 
the extent to which the gluon distribution dominates the valence distributions as one goes to small x. Indeed, 
even at $x\sim 0.1$ there are twice as many gluons as up and down valence quarks combined. Therefore at 
small $x$ we are constructing a theory where the dynamical degrees of freedom are gluons. The valence 
partons act as sources and the following simple kinematics suffices to convince one that they are static sources:
\begin{eqnarray}
\tau_{\rm wee} &=& {1\over k^-} = {2k^+\over k_\perp^2} \equiv {2xP^+\over k_\perp^2} \nonumber \\
\tau_{\rm valence} &\approx& {2P^+ \over k_\perp^2}\, =>\, \tau_{\rm wee} << \tau_{\rm valence}.
\label{eq:7}
\end{eqnarray}
Not much changes with the valence parton distributions over the time scales of interest for the 
dynamics of the wee partons. Thus one has something akin to the Born-Oppenheimer approximation-an 
important ingredient in an effective theory. However, one cannot integrate out the valence sources completely 
out of the theory. This is because they carry color charge (lots of it as we shall see) and therefore couple 
to the wee partons~\footnote{We note here that the phenomenon of ``limiting fragmentation" observed in rapidity 
distributions over a wide swath of energies and reactions is basically a statement about the recoil-less universal nature of the valence ("fragmentation") distributions.}. 

\begin{figure}[htbp]
\begin{center}
\resizebox*{!}{4.0cm}{\includegraphics{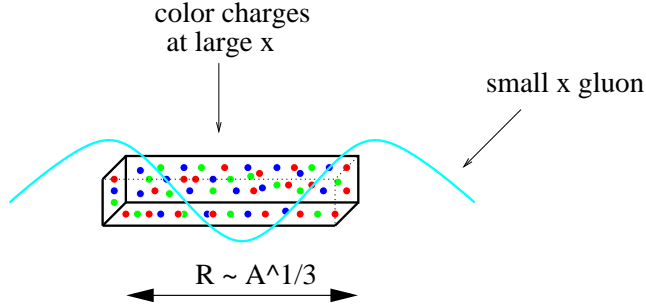}}
\end{center}
\caption{\label{fig:nine} A long wavelength wee parton  sees a lot of static charges ($\propto A^{1/3}$) when it resolves 
small transverse distances.}
\end{figure}

Since the wee parton has a large wavelength in the longitudinal direction, it can resolve a lot of color charges provided its transverse wavelength is not too large. The inequality 
\begin{equation}
\lambda_{\rm wee} \sim {1\over k^+} \equiv {1\over xP^+} >> \lambda_{\rm valence}\equiv {R\,m_p\over P^+}\,,
\label{eq:8}
\end{equation}
suggests that wee partons with $x<< A^{-1/3}$ can resolve partons all along the longitudinal extent of 
the nucleus. This is shown in Fig.~\ref{fig:nine}. Here $m_p$ is the nucleon mass
and $\gamma \sim P^+/m_N$ is the Lorentz factor in the infinite
momentum frame.
However, if the wee parton had a wavelength $k_\perp \leq \Lambda_{\rm QCD}\sim 1$ fm, it would see no color charge at all since color is confined (in nucleons!) on 
this scale. It is only if the wee parton has a short wavelength in the transverse direction $k_\perp >> \Lambda_{\rm QCD}$ that it will see color charges from different nucleons along the longitudinal direction. These charges 
will be random since they are confined to different nucleons and do not know about each other. 

How many of these random sources the wee partons actually couple to
depends on the typical transverse momentum of the wee
parton~\footnote{The wee parton is soft only in its longitudinal
momentum-its transverse momentum may be large.}. A wee parton with
momentum $p_\perp$ resolves an area in the transverse plane $
(\Delta x_\perp)^2 \sim 1/p_\perp^2$. The number of valence partons it
interacts
simultaneously with is then
\begin{equation}
k \equiv k_{(\Delta x_\perp)^2} = {N_{\rm valence}\over \pi R^2}\, (\Delta
x_\perp)^2
\, ,
\label{eq:9}
\end{equation}
which indeed is proportional to $A^{1/3}$ since $N_{\rm
valence}=3\cdot A$ in QCD.  

For a large nucleus with $k >>1$, one can show that the most likely representation is a higher dimensional  classical representation of order $\sqrt{k}$. The net charge that the wee partons couple to is a classical color 
charge and is represented by a classical color charge density (per unit transverse area) we denote as 
$\rho$. This argument can be made rigorous for $SU(N_c)$ for large numbers of random sources~\cite{SR}. 
Since the charges are random, we have
\begin{equation}
<\rho^a(x_\perp)> = 0\,\,;\,\,
<{\rho^a(x_\perp)\rho^b(y_\perp)}>
= \mu_A^2\, \delta^{ab}\,\delta^{(2)}(x_\perp -y_\perp) \, ,
\label{eq:10} 
\end{equation}
where 
\begin{equation}
\mu_A^2 = {g^2 A\over 2\pi R^2} \,\,,
\label{eq:11}
\end{equation}
is the color charge squared per unit area. For a very large nucleus, $A >> 1$, $\mu_A^2 \propto A^{1/3} >>
\Lambda_{\rm QCD}^2$ is a large scale. Since it is the only scale in the effective theory, we expect that 
$\alpha_S(\mu_A^2) << 1$. Thus in the large $A$ limit, one can construct the small x limit of QCD as a 
weakly coupled effective field theory.

After these preliminaries, we can now write down the generating functional for the small x effective action,
\begin{equation}
{\cal Z}[j] = \int [d\rho]\,W_{\Lambda^+}[\rho]\,\left\{{\int^{\Lambda^+}[dA]\delta(A^+)e^{iS[A,\rho]-\int j\cdot A}}
\over {\int^{\Lambda^+}[dA]\delta(A^+)e^{iS[A,\rho]}}\right\} \, ,
\label{eq:12}
\end{equation}
where $\Lambda^+$ is the longitudinal momentum scale separating the sources from the fields. $W_{\Lambda^+}[\rho]$ is a gauge invariant functional describing the distribution of sources at the 
scale $\Lambda^+$. The small x effective action can be written in terms of the sources and the fields as 
\begin{equation}
S[A,\rho] = {1\over 4}\int d^4 x\, F_{\mu\nu}^a \,F^{\mu\nu,a}
+ {i\over N_c}\int d^2 x_\perp dx^- \delta(x^-){\rm Tr} \left(\rho\, U_{-\infty,
\infty}[A^-]\right) \, ,
\label{eq:13}
\end{equation}
where 
\begin{equation}
U_{-\infty,\infty}={\cal P}\exp\left(ig \int dx^+ A^{-,a}T^a\right)
\label{eq:14}
\end{equation}
is a path ordered exponential along the light cone time direction. The 
second term in the effective action can also be written~\cite{JSR} as 
${\rm Tr}\left(\rho \ln(U_{-\infty,\infty})\right)$. To the order studied thus 
far, the two forms of this term, which represents the 
coupling between small x fields and large x sources, are equivalent. It remains an open question whether 
the latter form leads to different results at higher orders. 

As discussed previously, for a large nucleus, the sources are random light cone 
sources. For  $SU(N_c)$, this represents a random walk in the space of $N_c-1$ Casimirs. It can be shown explicitly~\cite{SR} 
that the quadratic Casimir dominates the random walk (with the higher Casimirs giving contributions that are suppressed 
by powers of $1/\sqrt{k}$. The weight functional in the path integral, as first conjectured in Refs.~\cite{MV,Kovchegov}, can therefore be represented as a Gaussian functional weight (for large $\sqrt{k}$) and one obtains,  
\begin{equation}
W[\rho] = \exp\left(-\int d^2 x_\perp {\rho^a\rho^a (x_\perp)\over 2 \mu_A^2}\right) \, ,
\label{eq:15}
\end{equation}
where $\mu_A^2$ was defined in Eq.~\ref{eq:11}. In general, $W[\rho]$ is not a Gaussian 
as we shall discuss shortly.

\begin{figure}[htbp]
\begin{center}
\resizebox*{!}{5.0cm}{\includegraphics{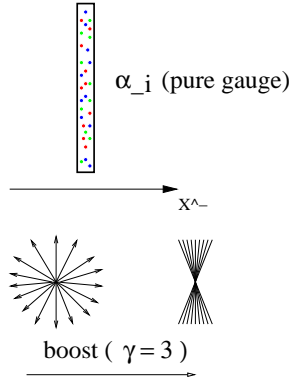}}
\end{center}
\caption{\label{fig:ten}Weizs\"{a}cker-Williams fields of a charge boosted to the infinite momentum frame are 
pure gauges on either side of the charge. }
\end{figure}

Before we go there, let us first discuss the effective action in Eq.~\ref{eq:13}. The classical equations of 
motion for a fixed configuration of $\rho$'s is given by the saddle point of the effective action in Eq.~\ref{eq:13}. 
These are just the Yang-Mills equations,
\begin{equation}
D_\mu F^{\mu\nu,a} = \delta^{\nu +}\,\delta(x^-)\,\rho^a(x_\perp) \, .
\label{eq:16}
\end{equation}
The solution of the Yang-Mills equations are the non-Abelian analog of the Weiz\"{a}cker--Williams 
fields in classical electrodynamics~\cite{MV}. As one may recall, when one boosts a classical charge 
in electrodynamics to the IMF, the fields of the charge look like that of a sheet of plane polarized radiation. 
The fields are singular on the sheet and pure gauges outside. This is shown in Fig.~\ref{fig:ten}

The solution is given by $A^-=0$ and 
\begin{equation}
A_{\rm cl.}^i = {1\over ig}{\cal P}\exp\left(ig{1\over \nabla_\perp^2}{\tilde \rho}(x_\perp,x^-)\right)\nabla_i
\exp\left(ig{1\over \nabla_\perp^2}{\tilde\rho}(x_\perp,x^-)\right)^\dagger \, .
\label{eq:17}
\end{equation}
The path ordering here is in $x^-$. Note also the $x^-$ dependence in ${\tilde \rho}$. A careful 
solution~\cite{JKMW,Kovchegov} of the Yang-Mills equations requires smearing in $x^-$. A final note on 
the solution is that the charge density ${\tilde \rho}$ that appears in the solution is {\it not} the color charge density in 
light cone gauge but instead the color charge density in covariant/Lorentz gauge. In the latter, one has the solution ${A^\prime}^+ = {1\over \nabla_\perp^2}{\tilde \rho}\,\delta(x^-)$, ${A^\prime}^- = {A^\prime}_\perp =0$. 

The explicit solution of the gauge field in terms of ${\tilde \rho}$ is not trivial. However, since we are interested in 
number distributions, one can simply replace the measure $[d\rho] \rightarrow [d{\tilde \rho}]$ in the path integral. 
The Jacobian in the transformation is quite simple~\cite{JKMW} and does not contribute to the result. Thus one 
can compute distributions in light cone gauge quite straightforwardly by expressing them in terms of charges in 
covariant gauge. To be specific, one averages the solution in Eq.~\ref{eq:17} 
with the weight functional $W$,
\begin{equation}
<AA>_\rho =\int [d{\tilde \rho}] A_{\rm cl.}[\rho]A_{\rm cl.}[{\tilde \rho}] W_{\Lambda^+}[{\tilde \rho}] \, .
\label{eq:18}
\end{equation}

\begin{figure}[htbp]
\begin{center}
\resizebox*{!}{5.0cm}{\includegraphics{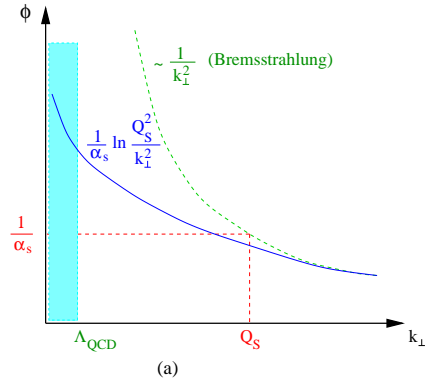}}
\end{center}
\caption{\label{fig:eleven} The occupation number of classical non-Abelian Weizs\"{a}cker-Williams 
fields.}
\end{figure}

For the Gaussian weight in Eq.~\ref{eq:15}, which is appropriate for large nuclei and not too small x, one 
can analytically compute the solution. The result is shown in Fig.~\ref{fig:eleven}. The occupation number 
is defined to be $\phi = (2\pi)^3 dN/\pi R^2/d^2 k_\perp dy/ 2(N_c^2-1)$ with $Q_s^2\approx 
\alpha_S N_c \mu_A^2 \ln(Q_s^2/\Lambda_{\rm QCD}^2)$. Hence $Q_s^2 \approx A^{1/3}\ln(A)
\sim A^{1/3}$ for $A>>1$. At large transverse momenta ($k_\perp >> Q_s$), 
the distributions have the characteristic 
$\phi\propto \mu_A^2/k_\perp^2$ form of Weizs\"{a}cker-Williams gluons in electrodynamics. At 
smaller transverse momenta ($k_\perp << Q_s$), the distribution has the form $\phi\sim 
\ln(Q_s/k_\perp)/\alpha_s$. Thus in light cone gauge, the strongly  non-linear behavior of the 
fields is responsible for the softening of the infrared behaviour of the classical fields. This
non-linearity is responsible for the phenomenon of saturation. A full description 
however requires a discussion beyond the classical level discussed thus far. 

We are now in a position to understand the term Color Glass Condensate for the state 
we are describing, namely, the ground state properties of a hadron/nucleus at very high 
energies. Color is obvious since the state is comprised of a large number of gluons. It is a 
condensate because, as we have just seen explicitly, the gluons have occupation numbers 
$\phi\sim 1/\alpha_S$ and have momenta peaked at $k_\perp\sim Q_s$. Finally, it is a 
glass because the gluons are coupled to random sources with time scales of evolution 
much longer than those of natural time scales associated with the scattering. 

\begin{figure}[htbp]
\begin{center}
\resizebox*{!}{2.54cm}{\includegraphics{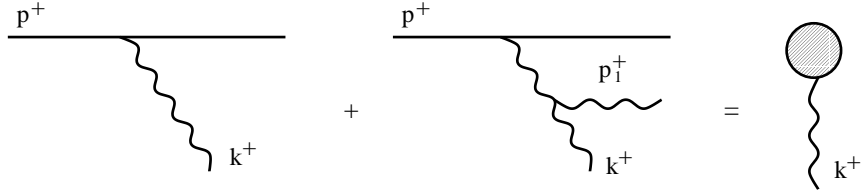}}
\end{center}
\caption{\label{fig:twelve}The weight functional modified under evolution. }
\end{figure}

We have discussed thus far a classical effective field theory for large nuclei and Gaussian 
sources. Though many features of the theory persist, the theory is inadequate to describe
 small x evolution at very small x. The main culprit is the Gaussian assumption for 
 $W[\rho]$. One does not obtain Gaussian correlations when quantum corrections are included. The 
 first suggestion this was the case came from computing small quantum fluctuations about the classical 
 saddle point solution. It was noticed that the corrections, proportional to $\alpha_S\,\ln(1/x)$, became 
 arbitrarily large at small x~\cite{AJMV}. It therefore was important to sum these up {\it a la} BFKL~\cite{JKMW}. 
The method that was invented to do this, now called by the acronym JIMWLK~\cite{JIMWLK} after the names of the 
principal contributors, was a Wilsonian renormalization group approach. Small fluctuation corrections to the effective action in Eq.~\ref{eq:13} at one step in $x$ ($\Lambda^+$) are incorporated into a modified 
source density at the next step: $W_{\Lambda^+}[\rho]\rightarrow W_{{\Lambda^\prime}^+}[\rho^\prime]$, 
where $\rho^\prime = \rho + \delta \rho$, is the new classical source density which incorporates the small 
fluctuations in the fields at the scale $\Lambda^+$ into the source density at the new lower momentum scale ${\Lambda^\prime}^+$. (One requires that $\alpha_S \ln(\Lambda^+/{\Lambda^\prime}^+)<<1$ to ensure that 
the small fluctuations are under control at each step.). These are shown in Fig.\ref{fig:twelve}.

The JIMWLK evolution equation for the RG evolution of the weight functional in Eq.~\ref{eq:12} can be 
written as
\begin{equation}
{\partial W_x[\rho]\over \partial \ln(1/x)} = {1\over 2}\,\int_{x_\perp,y_\perp} {\delta 
\over \delta \rho^a(x_\perp)}\chi^{ab}(x_\perp,y_\perp)[\rho]{\delta \over 
\delta \rho^b(y_\perp)}W_x[\rho]\, ,
\label{eq:19}
\end{equation}
where 
$\chi^{ab}(x_\perp,y_\perp)[\rho]=<\delta \rho^a(x_\perp)\delta\rho^b(y_\perp)>_{\rho}$ 
is a two point function in the background field of the hadron. From Eq.~\ref{eq:19}, one can 
construct a master equation for $n$-point correlators. With a change of variables, $\rho^a\rightarrow \alpha^a$ 
where $\nabla^2 \alpha = \rho$, one obtains for the expectation value $<O[\alpha]>_Y$ of an operator $O$ 
the relation, 
\begin{equation}
{\partial <O[\alpha]>_Y\over dY} = <{1\over 2} \int_{x_\perp,y_\perp} {\delta \over \delta \alpha^a(x_\perp)}
\chi^{ab}(x_\perp,y_\perp){\delta\over \delta \alpha^b(y_\perp)}O[\alpha]>_Y \, .
\label{eq:20}
\end{equation}
This equation is a generalized Fokker-Planck equation in functional space, where $Y$ is ``time" and $\chi$ is the 
diffusion coefficient~\cite{Weigert,IancuMcLerran}. $\chi$ has the functional form,
\begin{equation}
\chi_{x_\perp,y_\perp}^{ab}[\alpha] =\int {d^2 z \over 4\pi^3}\,\left\{(\vec{x}-\vec{z})\cdot (\vec{y}-\vec{z})\over 
(\vec{x}-\vec{z})^2(\vec{y}-\vec{z})^2\right\}\,\left\{(1-V_{x_\perp}^\dagger V_{z_\perp})(1-V_{z_\perp}^\dagger
V_{y_\perp})\right\}^{ab} \, ,
\label{eq:21}
\end{equation}
where 
\begin{equation}
V^\dagger(x_\perp) = {\cal P}\exp\left(ig\int dx^- \alpha^a (x^-,x_\perp) T^a\right) \, ,
\label{eq:22}
\end{equation}
and ${\cal P}$ denotes path ordering in $x^-$.

Consider for instance the two point function $<\alpha(x_\perp)
\alpha(y_\perp)>_Y$, we discussed previously. In the weak field limit, $g\alpha<<1$, the JIMWLK equation
for this two point function reduces to the BFKL equation~\cite{BFKL}. 

Interestingly, one can solve the JIMWLK equations for the two point functions in the other limit-the strong field limit-$g\alpha\sim 1$, 
using a mean field approximation, which we shall call the Random Phase Approximation (RPA)~\cite{IIM}. The phase space density $\phi$ we discussed in the 
classical approximation, has the following approximate solutions at large rapidities $Y$:
\begin{eqnarray}
\phi &\approx& {\mu_A^2\over k_\perp^2}\,\,\,\,{\rm for }\,\,\,\, \ln\left({k_\perp^2\over Q_s^2}\right) >> \alpha_S Y \,,
\nonumber \\
&\approx&  \left({\mu_A^2\over k_\perp^2}\right)^{1/2}\,e^{\omega {\bar \alpha}_s Y}\,\,\,\, {\rm for}\,\,\,\,
\ln\left({k_\perp^2\over Q_s^2}\right)\sim {\bar \alpha}_S Y\,\,\,\,{\rm but}\,\,\,\,k_\perp^2 >> Q_s^2(Y)\,,\nonumber \\
&\approx& {1\over \alpha_S}\,\ln\left({Q_s^2(Y)\over k_\perp^2}\right)\,\,\,\,{\rm for}\,\,\,\,k_\perp^2 << 
Q_s^2 \, .
\label{eq:23}
\end{eqnarray}
Here $\omega = 4 \ln 2$ is a constant and ${\bar \alpha}_S = \alpha_s N_c/\pi$. 
The phase space density in the first line (at very large $k_\perp$) corresponds to the MV/DGLAP regime discussed previously. The regime described in the second line here is new and is the BFKL regime of 
quantum evolution. The last line corresponds to the dense Color Glass regime of $k_\perp << Q_s$. 
Interestingly, the phase space density in this dense regime has the same structure as in the classical 
theory except that the color charges are screened at 
distances greater than $1/Q_s$~\cite{IIM2,Mueller3}, as opposed to $\Lambda_{\rm QCD}$ in the classical theory. A good approximation to the behavior of the phase space 
density in the three regions is the form~\cite{IIM2}
\begin{eqnarray}
\phi = {1\over \pi \gamma c {\bar \alpha_S}}\,\ln\left(1 + \left({Q_s^2\over k_\perp^2}\right)^\gamma\right) \, ,
\label{eq:24}
\end{eqnarray}  
where varying the anomalous dimension $\gamma$ from $0.63$ in the ``BFKL'' region to $1$ in the MV/DGLAP region. We will 
discuss this point further shortly.

The JIMWLK equations are master equations for $n$-point correlators under small x evolution. These 
do not have closed form expressions. For instance the evolution equation for 2-point correlators 
contains 4-point correlators, and so on. This hierarchy is analogous to the well known BBGKY hierarchy 
in statistical mechanics. Like the latter, one has to make assumptions about correlators at a certain order to 
close the hierarchy. The JIMWLK equations have not been solved analytically even though there has been 
an attempt to solve them numerically~\cite{RW}.

The Balitsky-Kovchegov equation~\cite{BK} is a simple non-linear evolution equation for the forward 
scattering amplitude. The derivation by Kovchegov was performed in the framework of Mueller's dipole 
picture~\cite{Mueller2}. It is equivalent to the JIMWLK expression for the forward scattering amplitude in 
the limit of large $N_c$ and $\alpha_S^2 A^{1/3} >> 1$. Recall the expression we had in Lecture I for 
the virtual photon-proton cross-section (Eq.~\ref{eq:3}). The hadronic (dipole) cross-section there can be written 
as~\cite{MV99}
\begin{equation}
\sigma_{q\bar q p}(x,r_\perp,b) = 2\int d^2 b \left(1- {\rm Re}\,S(x,r_\perp,b)\right) \, ,
\label{eq:25}
\end{equation}
where the $S$-matrix can be written in terms of the path ordered exponential in Eq.~\ref{eq:21} as
\begin{equation}
{\rm Re}\,S(x,r_\perp,b) = {1\over N_c}\,<{\rm Tr}\,(V^\dagger (x_\perp)V(y_\perp))>_Y \equiv 1-{\cal N}(x,r_\perp,b) \, ,
\label{eq:26}
\end{equation}
where ${\cal N}$ is the imaginary part of the forward scattering amplitude,  $\vec{r}_\perp=\vec{x}_\perp-\vec{y}_\perp$,  
$\vec{b} = (\vec{x}_\perp+\vec{y}_\perp)/2$ and the path ordered exponential $V$ was introduced in 
Eq.~\ref{eq:22}. Balitsky and JIMWLK have shown that the S-matrix in Eq.~\ref{eq:26} satisfies the equation 
\begin{equation}
{\partial <{\rm Tr}\,(V_x^\dagger V_y)>_Y\over \partial Y} = {{\bar \alpha}_S \over 2\pi}\,\int d^2 z\,{(x-y)^2 \over 
(x-z)^2 (z-y)^2}\, < {1\over N_c}\,{\rm Tr} (V_x^\dagger V_z) {\rm Tr}(V_z^\dagger V_y) - {\rm Tr}(V_x^\dagger V_y)> \, .
\label{eq:27}
\end{equation}
Here ${\bar \alpha_s} = \alpha_s N_c/\pi$. 

In the $N_c\rightarrow \infty$ and  $\alpha_S^2 A^{1/3}\rightarrow \infty$ limits, the average over the product of traces here can be written 
as the product of the averages of the traces~\footnote{One would imagine that in QCD that one would need only the 
large $N_c$ limit for this factorization. This can be seen for instance from the one loop effective action in QCD~\cite{Polyakov}. 
However, in our case, we are dealing with a very particular background field which does not factorize for large $N_c$ 
alone. It requires the large $A$ limit as well for 
factorization to hold. I thank F. Gelis for an enlightening discussion on this point.}
\begin{equation}
< {\rm Tr} (V_x^\dagger V_z) {\rm Tr}(V_z^\dagger V_y)> = <{\rm Tr} (V_x^\dagger V_z)>  <{\rm Tr}(V_z^\dagger V_y)> \, .
\label{eq:28}
\end{equation} 
With this factorization, one now obtains the Kovchegov ``mean field" equation for 
the imaginary part of the forward scattering amplitude
\begin{eqnarray}
{\partial {\cal N}(x_\perp,y_\perp)\over \partial Y}&=&{\bar\alpha_s}\int_{z_\perp}{(x_\perp-y_\perp)^2\over (x_\perp-z_\perp)^2(z_\perp-y_\perp)^2}\,
\Big[{\cal N}_Y(x_\perp,z_\perp)+{\cal N}_Y(y_\perp,z_\perp)-{\cal N}_Y(x_\perp,y_\perp) \nonumber \\
&-&{\cal N}_Y(x_\perp,z_\perp)\cdot {\cal N}_Y(z_\perp,y_\perp)\Big] \, .
\label{eq:29}
\end{eqnarray}

In the limit where ${\cal N}<<1$, the non-linear term in Eq.~\ref{eq:29} 
can be ignored and the equation reduces to the BFKL equation. In this limit, the amplitude has the solution, 
\begin{eqnarray}
{\cal N}_Y(r_\perp) &\approx& (r_\perp^2 Q_0^2)^{1/2}\,e^{\omega {\bar \alpha_S}Y}\exp\left(-{\ln^2(1/r_\perp^2 Q_0^2)\over 2\beta {\bar \alpha_S} Y}\right) \nonumber \\
&=& \exp\left( {\rho\over 2} + \omega {\bar \alpha}_S Y -{\rho^2\over 2\beta {\bar \alpha}_S Y} \right) \, ,
\label{eq:30}
\end{eqnarray}
where $\omega = 4 \ln 2\approx 2.77$, $\beta=28\zeta(3) \approx 33.67$ and $\rho = \ln(r^2 Q_0)^2$.  If we define the saturation condition as 
\begin{equation}
r_\perp \sim {2\over Q_s} \,\,=> \,\,{\cal N} = {1\over 2} \, ,
\label{eq:31}
\end{equation}
the vanishing of the exponent for this value of $r_\perp$ gives~\cite{IIM} 
\begin{equation}
Q_s^2 = Q_0^2 \,\, e^{c {\bar \alpha}_S Y} \,\,\,\,{\rm where}\,\,\,\, c=4.84 \, .
\label{eq:32}
\end{equation}
A more careful solution of the BK-equation close to the saturation boundary gives $c=4.88$. 

We can replace $Q_0$ in Eq.~\ref{eq:26} by $Q_s$ from Eq.~\ref{eq:32}. One thus obtains 
\begin{equation}
{\cal N}_Y(r_\perp) \approx (r_\perp^2 Q_s^2)^{\gamma_s}\,e^{\omega {\bar \alpha_s}Y}\exp\left(-{\ln^2(1/r_\perp^2Q_s^2)\over 2\beta {\bar \alpha_s} Y}\right) \, ,
\label{eq:33}
\end{equation}
with $\gamma_s\approx 0.63$. In the strong field RPA limit~\cite{LevinTuchin,IIM,IancuMueller}, one finds 
\begin{equation}
{\cal N}_Y = 1-{\kappa}\exp\left(-{1\over 4c}\,\ln^2(r_\perp^2 Q_s^2)\right) \, ,
\label{eq:34}
\end{equation}
with $c=4.84$ and $\kappa$ is an undetermined constant. A complete analytical solution of the BK equation 
is still lacking though, as we shall discuss, there have been interesting recent developments in this direction~\cite{MunierPeschanski,IMM,IT}. 

\begin{figure}[htbp]
\begin{center}
\resizebox*{!}{6.0cm}{\includegraphics{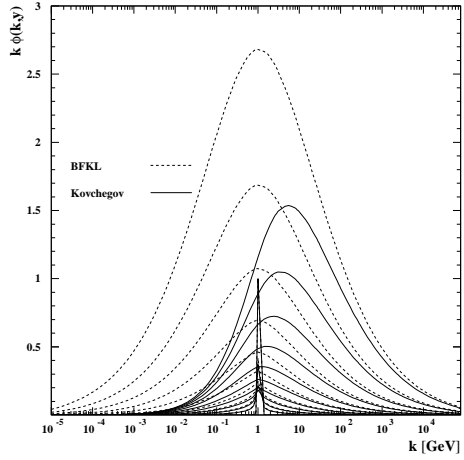}}
\end{center}
\caption{\label{fig:thirteen}The unintegrated gluon distribution $\phi$ from numerical solutions~\cite{GolecStastoMotyka} of the Balitsky-Kovchegov equation plotted as a function of transverse momentum. The dotted lines lines are the BFKL results 
while the solid ones are from solutions of the BK-equation.}
\end{figure}

\begin{figure}[htbp]
\begin{center}
\resizebox*{!}{6.0cm}{\includegraphics{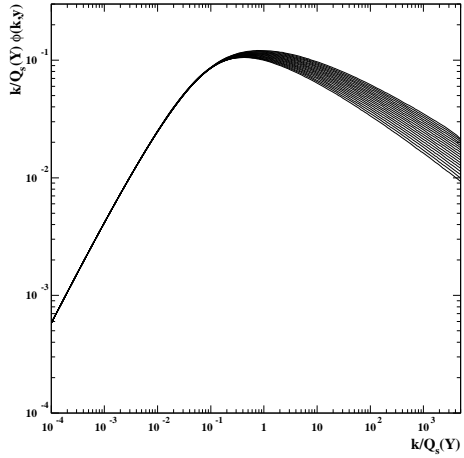}}
\end{center}
\caption{\label{fig:fourteen} Geometrical scaling of the unintegrated gluon distribution $\phi$ with $Q_s/k$ from 
numerical solutions~\cite{GolecStastoMotyka} of 
the Balitsky-Kovchegov equation.}
\end{figure}

The Balitsky-Kovchegov equation has however been solved numerically~\cite{Braun,ArmestoBraun,GolecStastoMotyka,Albacete}. These solutions have the following features:
\begin{itemize}
\item The dipole amplitude is shown to unitarize and solution is shown to exhibit geometrical scaling.
\item The saturation scale has the behavior predicted in Eq.~\ref{eq:32}.
\item The infrared diffusion problem of the BFKL solution is cured by the non-linear term in the 
BK equation. 
\item Ultraviolet diffusion is still present in the BK equation. This can be cured by including running 
coupling effects~\cite{ArmestoBraun,Albacete2}. 
\end{itemize}
Numerical solutions of the Balitsky-Kovchegov equation are shown in Figs.~\ref{fig:thirteen} and 
~\ref{fig:fourteen}. In Fig.~\ref{fig:thirteen}, the unintegrated gluon distribution is shown to be stable 
against infrared diffusion. The unintegrated gluon distribution from the BFKL equation is shown for 
comparison. In Fig.~\ref{fig:fourteen}, the solution is shown explicitly to exhibit geometrical scaling for 
$k_\perp \leq Q_s$. The solutions also exhibit a soliton-like structure; as shown by Munier and Peschanski, 
this analogy is an exact one~\cite{MunierPeschanski}. 
Numerical solutions to the full JIMWLK equations have been studied by Rummukainen and Weigert~\cite{RW}. 
Much work remains in that direction.

As we saw in Fig.~\ref{fig:five}, the HERA data appears to exhibit geometrical scaling up to 
rather large values of $Q^2$, values significantly larger than $Q_s^2$. Why is this so? A 
qualitative explanation~\footnote{The numerical solutions of course provide a quantitative answer, 
but don't necessarily provide insight into the phenomenon.} was provided by Iancu, Itakura and McLerran~\cite{IIM}. (See also Ref.~\cite{Mueller4}.) We can write the solution of the BFKL equation 
in Eq.~\ref{eq:30} as 
\begin{equation}
{\cal N}_Y(r_\perp) \approx \exp\left(\omega {\bar \alpha_s}Y-{\rho\over 2}-{\rho^2\over 2\beta{\bar \alpha_s}Y}
\right) \,,
\label{eq:35}
\end{equation}
where $\rho=\ln(1/r_\perp^2 Q_0^2)$. Now from our definition of the saturation scale (Eq.~\ref{eq:32}), we can write 
$\rho = \rho_c + {\bar \alpha}_S Y$, where $\rho_c = \ln(1/r_\perp^2 Q_s^2)$. One finds that the solution scales as long as 
$\rho_c^2 << 2\beta {\bar \alpha}_S Y$. Since we are interested in the region $Q^2 >> Q_s^2$, we find that geometric scaling 
holds for $\lambda {\bar \alpha_S} Y << \ln(Q^2/Q_0^2) << \lambda {\bar \alpha_S} Y + \sqrt{2\beta {\bar \alpha}_S Y}$. This is 
clearly  valid for $\sqrt{2\beta {\bar \alpha}_S Y} >> 1$ and is thus easily satisfied even for moderate $Y$'s.

A quantitative fit to HERA data using a dipole parametrization matching the BFKL dipole with the mean field (RPA) 
form in Eq.~\ref{eq:34} was performed by Iancu, Itakura and Munier~\cite{IIMunier}. They obtain good fits but 
the HERA inclusive data at present cannot distinguish between these fits and models that do not contain saturation~\cite{Forshaw}. In general, a large number of models give good fits to the inclusive data. Its the non-inclusive data that provides a more sensitive 
test and preliminary evidence there is that the CGC models fare better than those without saturation effects~\cite{Forshaw2}. To summarize, as discussed in the previous lecture, CGC based models do reasonably well but with the cautionary caveat that much more work remains for a consistent phenomenology.

\begin{figure}[htbp]
\begin{center}
\resizebox*{!}{5.0cm}{\includegraphics{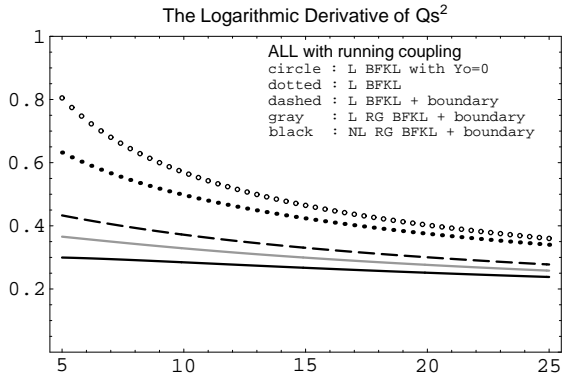}}
\end{center}
\caption{\label{fig:fifteen} The energy dependence of the saturation scale expressed in terms of $\lambda = d\ln Q_s^2/dY$.}
\end{figure}

How does $Q_s$ behave as a function of rapidity (or energy)? For fixed coupling, as we discussed previously, 
we obtain Eq.~\ref{eq:32}. If we allow the coupling to run, we get very simply, 
\begin{equation}
Q_{s,{\rm running}\,\alpha_s}^2 = \Lambda_{\rm QCD}^2 \exp\left(\sqrt{2b_0 c (Y+Y_0)}\right) \, ,
\label{eq:36}
\end{equation}
where $b_0$ is the coefficient of the logarithm in the one loop QCD $\beta$-function. The former expression can be understood as the 
low energy (rapidity) limit of the latter.  Now, the inclusion of 
running coupling effects in BFKL is highly non-trivial. The proper treatment of next-to-leading order BFKL 
requires renormalization group improvement of the NLO BFKL kernel, where all collinear singularities are 
summed to all orders~\cite{CCS}. It was suggested by Mueller and Triantafyllopolous that the saturation scale could be extracted 
from solving the BFKL equation in the presence of an absorptive boundary~\cite{Mueller4}. Triantafyllopolous 
showed~\cite{Dionysis} that this treatment could be extended to the resummed NLO BFKL case, and he was able to 
extract the energy dependence of $Q_s$. His result is shown in Fig.~\ref{fig:fifteen}. The value of $\lambda$ is rather large 
for the fixed coupling case. Furthermore, pre-asymptotic effects are 
clearly seen to be important.They are less so for running coupling. The ``naive"
running coupling result is seen to be a fairly good approximation to the full resummed result. The latter is seen to 
be to be quite close to the value extracted from the HERA experiments.

\begin{figure}[htbp]
\begin{center}
\resizebox*{!}{6.0cm}{\includegraphics{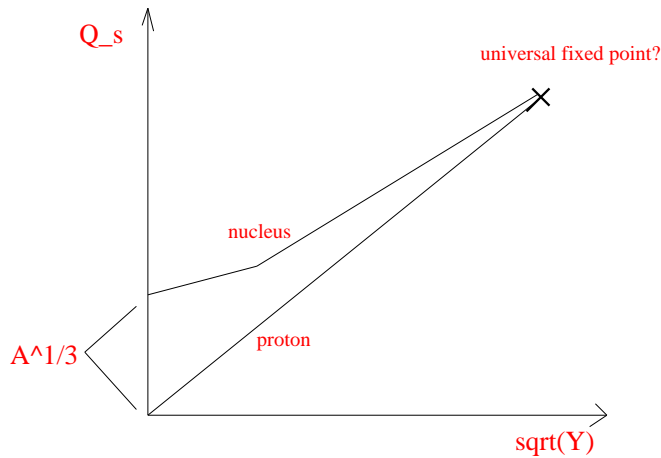}}
\end{center}
\caption{\label{fig:sixteen}The A-dependence of the saturation scale as a function of the square root of the rapidity. The initial relative 
height is of order $A^{1/3}$. }
\end{figure}

The A-dependence of the saturation scale $Q_s$ has been computed recently by Mueller in this 
picture~\cite{Mueller5}. The result is shown schematically in Fig.~\ref{fig:sixteen}.

This result has the form $Q_s^2 \approx \Lambda_{\rm QCD}^2\, e^{\sqrt{Y + \xi \ln^2(A^{1/3})}}$, 
where $\xi$ here is an undetermined 
constant. At small rapidities, the second term dominates and one has the result we discussed previously in lecture II; namely, $Q_s^2 \propto A^{1/3}$. For large $Y$, the converse is true, and the $A$ dependence of the 
saturation scale is gradually lost. Thus as illustrated in Fig.~\ref{fig:sixteen}, for large rapidities, one may find the 
remarkable result that the saturation scale, for fixed impact parameter, is independent of the target.

Finally, we now turn to very recent developments. As we mentioned previously, Munier and Peschanski made the remarkable observation 
that the Balitsky-Kovchegov equation, in a diffusion approximation (see footnote 3), can be shown to lie (with 
appropriate field re-definitions) in the same universality class as  the Fischer-Kolmogorov-Petrovsky-Piscunov (FKPP) equation~\cite{FKPP}, which describes the behavior of travelling wave fronts. Solutions to the latter are known, and Munier and Peschanski were able to deduce 
from these the asymptotic form of the amplitude in both the fixed and running coupling B-K equation as 
well as the form of the saturation scale. They were also able to compute the first two universal pre-asymptotic 
contributions (in rapidity) to the saturation scale. The first pre-asymptotic correction was also 
computed by Mueller and Triantafyllopoulos~\cite{Mueller4,Dionysis}, whose approach, 
as we discussed previously was apparently completely different. 

Thus far, we have not discussed the impact parameter dependence of the amplitudes. As we shall see, 
this ties into the previous discussion in an interesting way.
It was observed by Mueller and Shoshi~\cite{MuellerShoshi} that, for a fixed impact 
parameter, the amplitude ${\cal N}(b)$ could violate unitarity (in dense "hot spots"), even if the impact parameter averaged amplitudes remained well below unity. The possibility that this 
could happen in dipole-dipole scattering was suggested by Mueller and Salam~\cite{MuellerSalam} and observed in numerical simulations of the same by Salam~\cite{Salam}. Mueller and Shoshi suggested that 
these hot spot violations of unitarity could be corrected for by solving the BFKL equation with a second absorptive boundary~\footnote{The second boundary also had the virtue of correcting for the frame dependence of the BK-equation at high energies.} at ${\cal N}= \alpha_S^2$ in addition to the first one at ${\cal N}\approx 1$. 
Their computation of the saturation scale showed large corrections which vanished only at asymptotically 
small values of $\alpha_S$.

It was noticed subsequently by Iancu, Mueller and Munier~\cite{IMM} that these results could be interpreted 
in terms of a generalization of the FKPP equation, the stochastic FKPP (sFKPP) equation, well known to 
experts in statistical mechanics studying the properties of traveling wave fronts~\cite{Saarlos}. It was 
realized in these works that the original FKPP equation, in the region where the wavefront is small, did not 
properly treat the contribution of diffusion to the evolution of wavefronts.  It is this diffusive property which 
provides the seed for the evolution of  wavefronts in regimes where the occupation number is extremely small~\cite{BrunetDerrida}. This diffusion is modeled by a stochastic Langevin term in the KFPP equation. The picture described by this 
sFKPP equation is now that one has an ensemble of wavefronts, each  one stable, and saturation scales that 
are Gaussian distributed among the fronts. A very recent comprehensive analysis of the implications of these 
ideas for BK and JIMWLK has been provided by Iancu and Triantafyllopoulos~\cite{IT}. These are severe. While 
the physics in the saturation regime at high energies is likely well described by BK and JIMWLK, the competition 
between the dispersion of wavefronts (given by the variance of the Gaussian) and the scale governing the 
exponential decay of fronts (the BK anomalous dimension) suggests that there may be no BFKL regime 
at any energy. Iancu and Triantafyllopoulos have suggested a Langevin generalization of Balitsky's 
hierarchy which includes these effects. The ramifications of these ideas (particularly how they arise in 
the Color Glass Condensate) remain to be fleshed out further. However it should be clear to the reader that the stochastic nature of high energy processes and the exact analogies to well plumbed models in statistical physics is an exciting direction for research in high energy QCD.

\section{Lecture III}

We now come to the final topic of these lectures, high energy hadronic scattering.  Consider for instance the scattering of two 
nuclei at high energies-a problem very relevant at high energies. In the CGC picture, an observable $<{\cal O}>_Y$ can be computed as 
\begin{eqnarray}
<{\cal O}>_Y = \int [d\rho_1]\, [d\rho_2]\, W_{x_1}[\rho_1]\, W_{x_2}[\rho_2]\, {\cal O}(\rho_1,\rho_2) \, ,
\label{eq:37}
\end{eqnarray}
where $Y = \ln(1/x_F)$ and $x_F = x_1 - x_2$. All operators at small $x$ can be computed in the background 
classical field of the nucleus at small $x$. In particular, the Yang-Mills equations are solved to determine the 
dependence of the background field in terms of the sources~\footnote{For more detailed discussions of 
computing amplitudes in strong background fields, see Refs.~\cite{Baltz,GelisPeshier,FR,BFR1}.}. All quantum 
information, to leading logarithms in $x$, is contained in the source functionals $W_{x_1 (x_2)}[\rho_1(\rho_2)]$. 

To be specific, inclusive gluon production in the CGC is computed by solving the Yang-Mills equations 
$[D_\mu,F^{\mu\nu}]^a = J^{\nu,a}$, where 
\begin{eqnarray}
J^\nu = \rho_{p1}\delta(x^-)\delta^{\nu +} + \rho_{p2}\delta(x^+)\delta^{\nu -} \, .
\label{eq:38}
\end{eqnarray}
with initial conditions given by the Yang-Mills fields of the two nuclei before the collision. These are obtained 
self-consistently by matching the solutions of the Yang-Mills equations on the light cone. The initial conditions 
are determined by requiring that singular terms in the matching vanish. Since we have argued in the previous 
lectures that we can {\it compute} the Yang-Mills fields in the nuclei before the collision, the classical 
problem is in principle completely solvable. Similarly, quark pair production at small x can be computed by solving the Dirac equation (again by matching on the light cone) in the 
background field of the two nuclei with the latter obtained from solutions of the Yang-Mills equations. 
\vskip 0.1in
{\it This approach therefore solves the problem of the behavior of wee partons first outlined by Bjorken 
nearly 30 years ago~\cite{Bj1}. As we shall discuss in the following, this enables one to calculate from first 
principles the initial energy density and the formation time of partons in a high energy heavy ion collision 
(first estimated by Bjorken~\cite{Bj2}).} 
\vskip 0.1in
Hadronic scattering in the CGC is studied in practice through a systematic power counting in the density of sources in powers of $\rho_{1,2}/k_{\perp;1,2}^2$. This power counting in fact is more relevant at high 
energies than whether the incoming projectile is a hadron or a nucleus. In addition, 
one can begin to study the applicability of both collinear and $k_\perp$ factorization at small $x$ in this 
approach. We shall discuss here how to interpret gluon and quark production at high energies in hadronic collisions for the cases when both hadrons provide either  dilute or dense sources for the scattering as well as the asymmetrical case of one source being dilute and the other being dense. 
\vskip 0.1in
\subsection{\bf Gluon and quark production in the dilute/pp regime: ($\rho_{p1}/k_\perp^2\,\rho_{p2}/k_\perp^2 <<1$)}

The power counting here is applicable either to a proton at small x, or to a nucleus (whose parton density 
at high energies is enhanced by $A^{1/3}$) at large transverse momenta. The 
relevant quantity here is $Q_s$, which, as one may recall, is enhanced both for large $A$ and small $x$. 
So as long as $k_\perp >> Q_s >> \Lambda_{\rm QCD}$, one can consider the proton or nucleus as being dilute. 

\begin{figure}[htbp]
\begin{center}
\resizebox*{!}{4.0cm}{\includegraphics{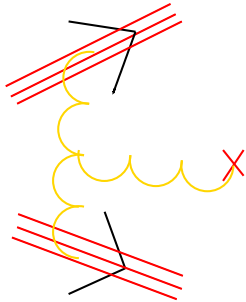}}
\end{center}
\caption{\label{fig:seventeen} $k_\perp$ factorized monojet.}
\end{figure}

To lowest order in $\rho_{p1}/k_\perp^2$ and $\rho_{p2}/k_\perp^2$, one can compute inclusive gluon production analytically. 
This was first done in the $A^\tau=0$ gauge ~\cite{KMW} and subsequently in the Lorentz gauge $\partial_\mu A^\mu=0$~\cite{KovRischke}.  At large transverse momenta, $Q_s << k_\perp$, the scattering can be expressed in a $k_\perp$-factorized 
 form shown in Fig.~\ref{fig:seventeen}. The inclusive cross-section can be expressed as the product of 
 two unintegrated ($k_\perp$ dependent) distributions times the matrix element for the scattering. This 
 type of factorization (as distinguished from the collinear factorization of perturbative QCD) is called 
 $k_\perp$-factorization. In this case, all the small x evolution can be factorized into the unintegrated 
 gluon distributions.

The comparison of this result to the collinear pQCD $gg\rightarrow gg$ process and the $k_\perp$ factorized $gg\rightarrow g$ was performed in Ref.~\cite{GyulassyMcLerran}.  At this order, the result is equivalent 
to the perturbative QCD result first derived by Gunion and Bertsch~\cite{GunionBertsch}. This 
result for gluon production is substantially modified, as we shall discuss shortly, by high parton density effects in the nuclei.

One can show that $k_\perp$ factorization is a good assumption at large momenta for quark pair-production 
as well. This was worked out in the CGC approach by Fran\c cois Gelis and myself~\cite{FR}.
The result for the gauge field obtained in Ref.~\cite{KovRischke} is put to use here. The result for inclusive quark pair production can  be expressed in $k_\perp$ factorized form as
\begin{eqnarray}
\frac{d\sigma_1}{dy_p dy_q d^2p_\perp d^2q_\perp}
&=&\frac{1}{(2\pi)^6 C_{_{A}}^2} 
\int\frac{d^2 k_{1\perp}}{(2\pi)^2}\frac{d^2 k_{2\perp}}{(2\pi)^2}
\delta(k_{1\perp}+k_{2\perp}-p_\perp-q_\perp)\,\nonumber \\
& & \varphi_1(k_{1\perp}) \varphi_2(k_{2\perp})
\frac{{\rm Tr}\,\left(\left|m^{-+}_{ab}(k_1,k_2;q,p)\right|^2\right)}
{k_{1\perp}^2 k_{2\perp}^2} \; ,
\label{eq:39}
\end{eqnarray}
where $\phi_1$ and $\phi_2$ are the unintegrated gluon distributions in the projectile and target respectively (with the gluon distribution defined as $xG(x,Q^2) = \int_0^{Q^2} d(k_\perp^2)\, \phi(x,k_\perp)$). The matrix element ${\rm Tr}\,\left(\left |m^{-+}_{ab}(k_1,k_2;q,p)\right|^2\right)$ is identical to the result derived in the $k_\perp$--factorization approach~\cite{CiafHaut,CollinsEllis}. This result has been applied extensively to study heavy quark production at collider energies~\cite{Shabelski}. In the limit 
$|\vec{k_{1\perp}}|\,,|\vec{k_{2\perp}}|\rightarrow 0$, $\frac{{\rm Tr}\,\left(\left|m^{-+}_{ab}(k_1,k_2;q,p)\right|^2\right)}
{k_{1\perp}^2 k_{2\perp}^2}$ is well defined--after integration over the azimuthal angles in Eq.~\ref{eq:39}, one obtains the usual matrix 
element $|{\cal M}|_{gg\rightarrow q\bar q}^2$, recovering the lowest order pQCD collinear factorization result.  
\vskip 0.1in
\subsection{\bf Gluon and quark production in the semi-dense/pA region ($\rho_{p}/k_\perp^2<<1\,\rho_{A}/k_\perp^2\sim 1$).}

The power counting here is best applicable to asymmetric systems such as proton-nucleus collisions, which 
naturally satisfies the power counting for a wide range of energies. Of course, as one goes to extremely high 
energies, it is conceivable that the parton density locally in the proton can become comparable to that in the 
nucleus. We shall discuss how that case works as well~\footnote{The reader might wonder about the 
frame dependence in this formulation. Our formalism is valid as long as the infinite momentum frame is 
applicable for {\it both} projectile and target, namely their momenta are much larger than the momenta of 
the constituents in the hard scattering.}.

In the semi-dense/pA case, one again solves the Yang--Mills equations $[D_\mu,F^{\mu\nu}]=J^\nu$, now with the light cone sources 
$J^{\nu,a}$ (=$\delta^{\nu+}\,\delta(x^-)\,\rho_p^a(x_\perp)+\delta^{\nu -}\,\delta(x^+)\,\rho_A^a(x_\perp)$),
to determine the gluon field produced-to lowest order in the proton source density and to all orders in the  
nuclear source density. The computations are performed in Lorentz/covariant gauge $\partial_mu A^\mu=0$. The inclusive gluon production cross-section, in this framework, was first computed by Kovchegov and Mueller~\cite{KovMueller} and shown to be $k_\perp$ factorizable in Ref.~\cite{KKT}. In Ref.~\cite{BFR1}, the gluon field produced in pA collisions was computed 
explicitly in Lorentz gauge. One obtains, 
\begin{eqnarray}
{ A}^\mu(q)&=&{ A}_{p}^\mu(q)
+\frac{ig}{q^2+iq^+\epsilon}
\int\frac{d^2k_{1\perp}}{(2\pi)^2}
\Big\{
C_{_{U}}^\mu(q,k_{1\perp})\, 
\big[U(k_{2\perp})-(2\pi)^2\delta(k_{2\perp})\big]
\nonumber\\
&&\qquad\qquad
+
C_{_{V}}^\mu(q)\, 
\big[V(k_{2\perp})-(2\pi)^2\delta(k_{2\perp})\big]
\Big\}
\frac{{ \rho}_p(k_{1\perp})}{k_{1\perp}^2}
\; ,
\label{eq:40}
\end{eqnarray}
with $k_2=q-k_1$ and $U$ \& $V$ are path ordered Wilson lines containing all orders in the nuclear source density $\rho_A$.  The coefficient functions $C_U$ and $C_V$ are simply related to the well known Lipatov effective vertex  $C_{_{L}}^\mu$ through the relation $C_{_{L}}^\mu=C_{_{U}}^\mu+\frac{1}{2}C_{_{V}}^\mu$. 

The path ordered exponentials $U$ are color matrices arising from the rotation of the color charge density of the proton source due to multiple scattering off the nucleus.  The path ordered exponentials $V$ (differing from the $U$'s by a 1/2 factor in the argument of the exponential) arise from the Green's function solutions of the equations of motion. Interestingly, they do not appear in the final result for gluon production. This is because for gluons produced on shell one finds remarkably that $C_U\cdot C_V=C_V^2=0$ and $C_U^2=C_L^2 = 
4 k_{1\perp}^2 k_{2\perp}^2/q_\perp^2$. Thus only bi-linears of the Wilson line $U$  survive in the squared amplitude that gives us the gluon production cross-section. The result is $k_\perp$-factorizable 
analogous to Eq.~\ref{eq:36}, except now one replaces $\phi_2$ with the unintegrated nuclear gluon distribution $\phi_A\propto <U^\dagger U>_{\rho_A}$. This distribution contains powers of  the usual unintegrated gluon distribution to all orders-one recovers the usual unintegrated gluon distribution ($\varphi_2$ in Eq.~\ref{eq:39}) at large transverse momentum. 

Our result in Lorentz gauge is exactly equivalent to that of Dumitru \& McLerran in $A^\tau=0$ gauge~\cite{DumitruMclerran}. We now turn to a discussion of the Cronin effect which can be interpreted 
in this framework.

\vskip 0.1in
{\it The Cronin effect}

The Cronin effect was discovered in proton-nucleus collisions in the
late 70's~\cite{AntreCFSK1,KlubePSAC1,CroniFSBM1}. The effect observed
was a hardening of the transverse momentum spectrum in proton-nucleus
collisions, relative to proton-proton collisions, that sets in at
transverse momenta of order $k_\perp \sim 1-2$~GeV, and disappears at
much larger $k_\perp$'s. A corresponding depletion was seen at low
transverse momenta, accompanied by a softening of the spectrum. At
that time, the effect was interpreted as arising from the multiple
scatterings of partons from the proton off partons from the nucleus
\cite{KrzywEPS1}.  At high $k_\perp$, the higher twist effects, which, in
the language of perturbative QCD, are responsible for multiple
scattering~\cite{QiuS1,QiuS2} are suppressed by powers of $k_\perp$.
The relative enhancement of the cross-sections at moderate $k_\perp$'s
should thus die away -- and indeed, the data seemed to suggest as
much.  Though a qualitative understanding of the previously observed
Cronin effect was suggested by perturbative QCD, a quantitative
agreement for all its features (such as, for instance, the flavor
dependence) is still lacking. The Cronin effect has been observed at
central rapidity at RHIC, and has been interpreted in terms of
multiple scatterings, both in collinear factorized pQCD
\cite{KopelNST1,Vitev1,Accar1,AccarG2,AccarG3} and in saturation
inspired models
\cite{DumitruJamal,FrancoisJamal,BKW,JYR,Albacete}.

\begin{figure}[htbp]
\begin{center}
\resizebox*{!}{4.0cm}{\includegraphics{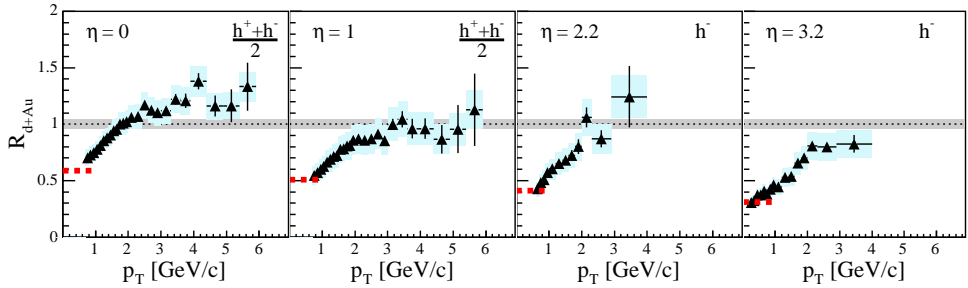}}
\end{center}
\caption{\label{fig:eighteen} Depletion of the Cronin peak from $\eta=0$ to $\eta=3$ for minimum bias events. From Ref.~\cite{Debbe1}.}
\end{figure}

\begin{figure}[htbp]
\begin{center}
\resizebox*{!}{4.0cm}{\includegraphics{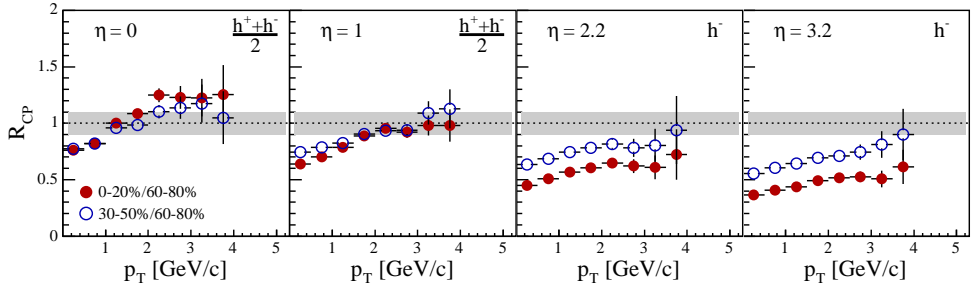}}
\end{center}
\caption{\label{fig:nineteen} Centrality dependence of the Cronin ratio as a function of rapidity. From Ref.~\cite{Debbe1}.}
\end{figure}

First data from RHIC on forward D-Au scattering at $\sqrt{s} = 200$ GeV/nucleon demonstrate how the Cronin effect is
modified with energy or, equivalently, with the rapidity. The $x$
values in nuclei probed in these experiments, at $k_\perp \sim 2$~GeV, range
from $10^{-2}$ in the central rapidity region down to $10^{-4}$ at
very forward rapidities~\footnote{Guzey, Strikman and Vogelsang have recently argued that the average $x$ of the BRAHMS data is significantly higher~\cite{GSV}. This is indeed true for the $2\rightarrow 2$ process. However, at small $x$, 
the $2\rightarrow 1$ process becomes important and samples smaller $x$'s in the nucleus than in the $2\rightarrow 2$ 
process.}. A dramatic result obtained by the 
BRAHMS experiment~\cite{Debbe1} which has taken data up to
pseudo-rapidities $\eta=3.2$~\footnote{The trends seen by BRAHMS are
also corroborated by PHOBOS, PHENIX and STAR in different kinematic
ranges~\cite{PHENIX1,STAR1,PHOBOS1}.} is the shrinking of the Cronin peak 
rapidly with rapidity and at higher rapidity, one sees in Fig.~\ref{fig:eighteen} that there is a
significant suppression instead. Equally interesting is the centrality
dependence of the effect.  At central rapidities, one
observes that the Cronin peak is enhanced in more central collisions,
while, for forward rapidities, the trend is reversed: more central
collisions at forward rapidities show a greater suppression than
less central collisions! This is shown in Fig.~\ref{fig:nineteen}.
 
These results have a natural qualitative understanding in the CGC framework. 
Comparisons of saturation inspired models to the RHIC D-Au data 
lead to the following key conclusions:
\begin{itemize}
  
\item Tree level partonic re-scattering responsible for the classical
Cronin effect in high energy Deuteron-Gold collisions is computable in
the CGC approach. When the weight for the color
sources is Gaussian (which is valid at moderately small x for large nuclei), the formalism reduces to the familiar Glauber
formalism of independent multiple scatterings. In the 
McLerran-Venugopalan model, a peak appears at $k_\perp\sim Q_{s,_A}$,
which is more pronounced for more central collisions~\cite{DumitruJamal,FrancoisJamal,BKW,JYR}.
  
\item Quantum corrections (due to small-$x$ evolution) to the tree
level partonic re-scattering can be ``naively'' included by letting
$Q_s\rightarrow Q_s(x)$. This model of
quantum evolution  predicts a larger Cronin effect at larger rapidities (because $Q_s$ grows as $x$ decreases) sharply disagreeing with the RHIC d-Au data at forward rapidities.  On theoretical grounds it has been known for some time
that quantum evolution quickly destroys~\footnote{This also explains why at sufficiently small $x$ it would be more consistent to use something like the BFKL+RPA fit of Ref.~\cite{IIMunier} in interpreting DIS data than Golec-Biernat-Wusthoff type fits.} the Gaussian weight functional~\cite{AJMV}.
  
\item Proper quantum evolution in the CGC is described by the JIMWLK
evolution equation. All simplified versions of JIMWLK (be they BK or RPA) lead to 
a ``Cronin suppression" at large rapidities~\cite{KLM,KKT,BKW,Albacete,IIT}. Furthermore, the anomalous 
dimensions governing the hardness of the momentum spectrum are, as discussed previously, of 
BFKL-type. These lead to a softer spectrum whose magnitude is more severely suppressed for more central 
collisions, exactly as predicted by the data. Another corroborative piece of evidence is the preliminary observation of 
the broadening of azimuthal correlations between forward going and central hadrons in D-Au collisions by STAR~\cite{STAR2}. This observation confirms a prediction by Kharzeev, Levin and McLerran~\cite{KLM2}.

\end{itemize}

From the qualitative features of the RHIC d-Au data the following picture
emerges~\footnote{This picture will have to be substantiated by more
detailed (read quantitative) computations and a thorough confrontation
to the experimental data.}. The MV model works reasonably well at
$x\sim 10^{-2}$ -- central rapidities at RHIC -- where small $x$ evolution 
is not significant. It is a good model of the
initial conditions for quantum evolution as a function of rapidity.
Correspondingly, the fact that the MV model fails badly at forward
rapidities suggests that quantum evolution effects cannot be
accounted for simply by a rescaling of the saturation momentum.  For 
instance, calculations with the RPA model  reveal a
qualitatively different behavior of the ratio $R_{dA}$. In particular,
a prediction of this mean-field solution is that the ratio $R_{pA}$
tends to $A^{(\gamma-1)/3}$ at large transverse momentum, where
$\gamma$ is the BK anomalous dimension. 

An important test of the CGC in D-Au collisions is 
provided by electromagnetic probes~\cite{FrancoisJamal,Jamal,BMS,Betemps}. If one observes the Cronin 
effect and its subsequent depletion with rapidity, one cannot but conclude that it is an initial state effect. Else, it 
may be predominantly due to final state effects, as has been proposed recently~\cite{FriesHwa}.

\vskip 0.1in
{\it Quark production in p/D-A collisions}

Quark production can now be computed with the gauge field in Eq.~\ref{eq:40}~\cite{BFR2}. The field is decomposed into the sum of `regular' terms and 'singular' terms; the latter contain $\delta(x^+)$. The regular terms are the cases where a) a gluon from the proton interacts with the nucleus and produces a $q\bar q$-pair outside, b) the gluon produces the pair which then scatters off the 
nucleus. Naively, these would appear to be the only possibilities 
in the high energy limit where the nucleus is a Lorentz contracted 
pancake. However, in the Lorentz gauge, one has terms identified with the singular terms in the gauge field which correspond to the case where the quark pair is both produced and re-scatters in the 
nucleus! Indeed, the contribution of this term to the 
amplitude cancels the contribution of the term proportional to the $V$'s (see Eq.~\ref{eq:40}) in the regular terms. 

Our result for quark pair production~\footnote{See also related work in Refs.~\cite{ShafNik} and ~\cite{Tuchin,KopRauf,Raufeisen} and for a recent review of $k_\perp$ factorization in heavy quark production, see 
Ref.~\cite{GoncalvesMachado}.}, unlike gluon production, is not strictly $k_\perp$ factorizable. It can however still be written in $k_\perp$ factorized form as a product of the unintegrated gluon distribution 
in the proton times a sum of terms with three unintegrated distributions,  $\phi_{g,g}$, $\phi_{q\bar q, g}$ and $\phi_{q\bar q,q\bar q}$. These are respectively proportional to 2-point, 3-point and 4-point correlators of the Wilson lines we discussed previously. For instance, the distribution $\phi_{q\bar q, g}$ can be interpreted as the probability of having a $q\bar q$ pair in the amplitude and a gluon in the complex conjugate amplitude. For large transverse momenta 
or large mass pairs, the 3-point and 4-point distributions collapse to the unintegrated gluon distribution, and we recover the previously discussed $k_\perp$-factorized result for pair production (Eq.~\ref{eq:39}) in the dilute/pp-limit. Due to the 
length of the expressions, we will not present them here but refer the reader to Ref.~\cite{BFR2}.

Single quark distributions are straightforwardly obtained. Here the 4-point correlator $\phi_{q\bar q,q\bar q}$ collapses upon 
integration over the momentum of the quark or anti-quark to the 2-point correlator $\phi_{q,q}$ corresponding to a quark (or 
anti-quark) in the amplitude and complex conjugate amplitude. 

For Gaussian sources, as in the MV-model, these 2-,3- and 4-point 
functions can be computed exactly as discussed in Ref.~\cite{BFR2}. Single quark distributions in the MV-model 
were recently computed by Tuchin~\cite{Tuchin}. Explicit computations of the size of $k_\perp$-violating 
terms as a function of quark mass and $Q_s$ are underway and will be reported shortly~\cite{FFR}. 

\vskip 0.1in
{\it The results for gluon and quark production in Proton/Deuteron-Gold collisions, coupled with the previous results for 
inclusive and diffractive~\cite{FrancoisJamal2,Kopeliovich,KT,KovWied} distributions in DIS suggest an important new paradigm. As one goes to 
smaller values of $x$ in DIS and hadron colliders, previously interesting observables such as gluon distributions are 
no longer the right observables to capture the relevant physics. Instead they should be replaced by these dipole and 
multipole correlators of Wilson lines that seem ubiquitous in all high energy processes and are similarly gauge invariant 
and process independent. The renormalization group running of these operators may be as powerful and sensitive a harbinger of new physics as were the parton distributions in the mid-70's.}

\subsection{\bf Gluon and quark production in the dense/AA region ($\rho_{A1}/k_\perp^2 =\rho_{A2}/k_\perp^2\sim 1$).}

Since $\rho_{1,2}/k_\perp^2\sim 1$, one has thus far not been able to compute particle production analytically in the 
CGC.  The problem however is well defined in weak coupling and can be solved numerically~\cite{AR,AYR,Lappi}.
In practice, this is equivalent to solving the Yang-Mills equations after the collision of two ultrarelativistic nuclei with 
initial conditions given by the classical gluon fields of each of the nuclei before the collision~\cite{KMW}. 

\begin{figure}[htbp]
\begin{center}
\resizebox*{!}{4.0cm}{\includegraphics{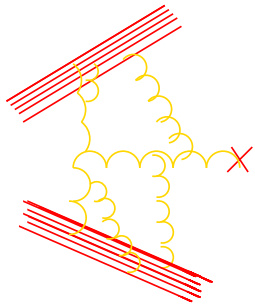}}
\end{center}
\caption{\label{fig:twenty}Diagrams which break $k_\perp$ factorization are $O(1)$ for $k_\perp\leq Q_s$-they are {\it not} suppressed.}
\end{figure}

The numerical simulations performed thus far assume Gaussian initial conditions as in the MV model. As we noted in 
our discussion of Deuteron-Gold collisions, these may actually be good initial conditions for central Gold-Gold collisions 
at RHIC where the typical $x$ is of order $10^{-2}$. These will not be good initial conditions at the LHC where the 
typical $x$ at central rapidities will be at least an order of magnitude lower. In that case, one has to use solutions of 
JIMWLK. 

{\it Unlike gluon production in the pp and pA cases, $k_\perp$-factorization breaks down in the AA-case~\cite{AR,Balitsky2}. $k_\perp$ factorization breaking diagrams of the sort shown in Fig.~\ref{fig:twenty} are of order 1 and there 
are a large number of these. A significant consequence is that one cannot factor the quantum evolution of the initial 
wavefunctions into unintegrated gluon distributions unlike the pA case.}

 Nevertheless, there is a systematic way to 
include small x effects in the AA case. Numerical solutions of the JIMWLK equations now exist and one can study the 
rapidity evolution of the Wilson lines we discussed in Lecture II~\cite{RW}. The numerical lattice formalism developed in Ref.~\cite{AR} is ideally suited to compute particle production in the forward light cone by matching the Wilson lines from 
each of the nuclei on the light cone. This program has not yet been carried out.

In the following, we will restrict ourself to discussing numerical solutions with Gaussian initial conditions. The saturation 
scale $Q_s$ (which is an input in the numerical solutions in this approximation) and the nuclear radius $R$ are the 
only parameters in the problem~\footnote{Another scale is the nucleon size-the color charge over this scale is constrained 
to be zero. Our results are insensitive to reasonable variations in this scale.}.
The number and energy of gluons released in a heavy ion collision of identical nuclei can therefore be simply expressed as
\begin{eqnarray}
{1\over \pi R^2}\,{dE\over d\eta} &=& {c_E\over g^2}\,Q_s^3 \,,\nonumber \\
{1\over \pi R^2}\,{dN\over d\eta} &=& {c_N\over g^2}\, Q_s^2 \, ,
\label{eq:41}
\end{eqnarray}
where (up to $10\%$ statistical uncertainity) $c_E=0.25$ and $c_N=0.3$. Here $\eta$ is the space-time 
rapidity. The numerical computations are performed in terms of the color charge 
squared per unit area $\mu_A^2$ that we discussed in lecture II (see Eq.~\ref{eq:11}). The results for $c_E$ 
and $c_N$ are obtained for these values. When re-expressed 
in terms of $Q_s$, there is a small logarithmic uncertainity depending on the scale of $Q_s$. 

\begin{figure}[htbp]
\begin{center}
\resizebox*{!}{4.0cm}{\includegraphics{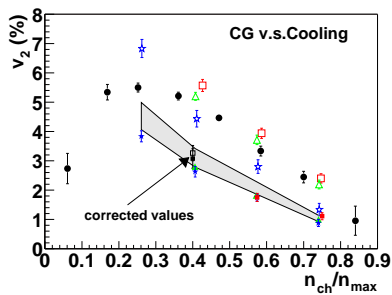}}
\end{center}
\caption{\label{fig:21} $v_2$ from melting colored glass.}
\end{figure}

The number distributions of gluons can also be computed in this approach. Remarkably, one finds that a) the 
number distribution is infrared finite, and b) the distribution 
is well fit by a massive Bose-Einstein distribution for $k_\perp/Q_s < 1.5$ GeV with a ``temperature" of $\sim 0.47 Q_s$ 
and by the perturbative distribution $Q_s^4/k_\perp^4$ for $k_\perp/Q_s > 1.5$. 

The RHIC data on the multiplicity (approximately 1000 hadrons in one unit of rapidity) and transverse energy 
(approximately $500$ GeV for central rapidities) of produced hadrons combined with Eq.~\ref{eq:41} place strong constraints on what $Q_s$ can be.  If $Q_s$ is too small, we will find, absurdly, that the initial transverse energy is 
less than the final measured transverse energy. On the other hand, if $Q_s$ is too large, we will find that 
the initial multiplicity of gluons is greater than the final multiplicity of hadrons. Besides, the initial 
energy per particle will be too difficult to get rid of to match the oberved experimental value. While there is no apparent 
theorem that prohibits the initial gluon multiplicity being greater than the final hadron multiplicity, such a situation is unlikely in all statistical/hydrodynamic scenarios of the 
RHIC collisions. These constraints therefore allow us to place the bound that~\cite{AYR2}. 
\begin{equation}
1.3 < Q_s < 2\,\, {\rm GeV}
\label{eq:42}
\end{equation}
This bound is consistent with an extrapolation of the Golec-Biernat--Wusthoff (see Lecture I) fit of $Q_s^2$ 
to the RHIC data (see Eq.~\ref{eq:5}). A simple extrapolation gives $Q_s\approx 1.4$ GeV.  Other considerations 
from RHIC will suggest (as we will see shortly) that the value of $Q_s$ should be closer to the higher end of this bound.

\begin{figure}[htbp]
\begin{center}
\resizebox*{!}{6.0cm}{\includegraphics{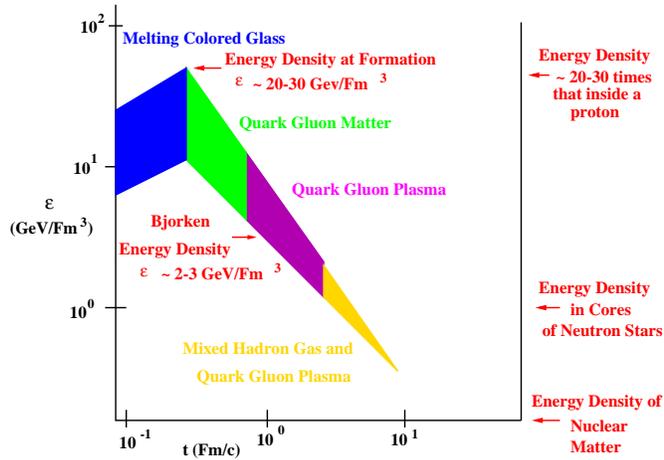}}
\end{center}
\caption{\label{fig:22}The space-time evolution of a RHIC collision. }
\end{figure}
In Ref.~\cite{AYR3}, we computed the elliptic flow produced by the CGC right after the collision. We found 
(see Fig.~\ref{fig:21}) that only about half the elliptic flow could be generated in this way. Besides, the 
$p_\perp$ dependence of the $v_2(p_\perp)$ distribution was much steeper than the RHIC data. This 
result strongly suggested that final state interactions were important at RHIC and that the RHIC data could 
not be explained by initial state interactions alone. If there are significant final state interactions and 
hydrodynamic flow (as suggested by successful hydrodynamic fits to the RHIC data), this suggests that $Q_s$ 
must be closer to the $2$ GeV upper bound. Otherwise, if it were instead closer to the lower bound in Eq.~\ref{eq:41}, 
the partonic $E_\perp$ at $\eta=0$ would be uncomfortably close to the observed hadronic value. Since $E_T/N \sim 0.9 \,Q_s$, for $Q_s\sim 2$ GeV, the 
system will have to do a significant amount of work to reduce the transverse energy per particle to the 
observed value; this is what one expects from hydrodynamics. Thus from very simple considerations, and from 
the RHIC data, we have already learned quite a bit about the initial state of the matter produced at RHIC. 
The initial energy density (at a time $\tau \sim 3/Q_s$) can be estimated to be between $20-40$ GeV/fm$^3$, much greater than the energy density required by the lattice to form a Quark Gluon Plasma (QGP). 

What is not understood is the transition to the QGP from the CGC. (A cartoon showing the timeline of a RHIC collision 
is shown in Fig.~\ref{fig:22}.) Due to the rapid expansion of the system, the occupation number of modes falls well below one on time scales of order $1/Q_s$. From these times onwards, one expects the 
canonical classical approach to break down-well before thermalization.  On the other hand, for elliptic flow from hydrodynamics to be significant, the conventional wisdom is that thermalization should set in early. A necessary condition is that momentum distributions should be isotropic. The CGC initial conditions are very anisotropic with $<p_\perp> \sim Q_s$ and $<p_z>\sim 0$. How does this isotropization take place? All estimates of final state re-scattering of partons formed 
from the melting CGC, both from $2\rightarrow 2$ 
processes~\cite{Mueller6,BV,SerreauSchiff} and $2\rightarrow 3$ processes~\cite{BMSS} suggest thermalization takes 
longer than what the RHIC collisions seem to suggest-in the latter case, $\tau_{\rm thermal}\sim {1\over \alpha_S^{13/5}}\,
{1\over Q_s}$, which at RHIC energies gives $\tau_{\rm thermal}\sim 2-3$ fm~\cite{BMSS2}. 

Recently, it has been suggested that collective instabilities, the non-Abelian analog of the well known Weibel instabilities 
in plasma physics, can speed up themalization~\cite{Stan,AML,RS1}. Starting from very anisotropic (CGC-like) initial conditions, these instabilities drive the system to isotropy on very short time scales, of order $1/Q_s$ in some estimates.

What is the relation of this language of instabilities and that of our classical field simulations? Why didn't we see such 
instabilities? One possibility is that our particular initial conditions, the non-linearities of the fields and the rapid expansion of the system kill the growth of instabilities. Another intriguing possibility is the following. In our solutions of the Yang-Mills equations, it is assumed that the incoming nuclei are delta-function sources 
on the light cone. This assumption directly leads to an explicit boost invariance of the classical fields produced in the 
collision. If one relaxes the delta function criterion, small violations of boost invariance should exist. Are these the seeds 
for the instabilities? Further, to properly study thermalization, one should better understand the interaction of high momentum 
(particle) and low momentum (field) degrees of freedom and their evolution. This leads to a real time renormalization group 
description~\cite{Boyanovsky} which should be coupled with the stability analysis. Work in these directions is ongoing 
and will hopefully soon provide further insight into the nature of thermalization in heavy ion collisions.

Much of the discussion about equilibration has focused on kinetic equilibration. However, equally interesting is the 
problem of chemical equilibration. At high energies, the initial state in a heavy ion collision is dominated by gluons. 
Sea quark pairs are produced from the gluon fields. Are they produced in sufficient numbers and do they re-interact 
sufficiently strongly for the system to reach chemical equilibrium (where the ratio of gluons to quarks is expected to 
be $32/21 N_f$)? Under normal circumstances, one would expect, in weak coupling, that the production of quarks to 
be suppressed. However, since the fields from the CGC are of order $1/g$, it is conceivable that the strong fields could 
drive the system to equilibrium. First steps have been taken to study this problem~\cite{Dietrich,FTK}. The problem involves 
solving the Dirac equation in the background field of the two nuclei (which, as we discussed previously, can be computed numerically). Hopefully further progress can be made on this problem in the near future.

Thus far, we have said very little about heavy ion phenomenology at RHIC. There is a successful CGC model-
the KLN model~\cite{KLN}-which does a very good job of explaining certain bulk features of the data such as the 
centrality dependence  and energy dependence of the multiplicity and the rapidity distributions at different RHIC energies. 
The model assumes $k_\perp$-factorization and parton-hadron duality. With regard to the former, as we have discussed, 
we expect $k_\perp$-factorization to be broken in AA-collisions. Thus the fact that KLN does well suggests perhaps that 
the effect is a small one, or that several factors contribute to the successful predictions. For instance, some effect of the 
factorization breaking might be taken into account by the `gluon liberation' factor (our $c_N$ in Eq.~\ref{eq:41}). What KLN has trouble with is the energy dependence-on face value one gets $E_\perp/N \sim Q_s$, about a factor of 3 larger 
than the RHIC data. This requires that post CGC either the system does a lot of isentropic work or partons multiply furiously or some combination of the two. Further, the RHIC data on elliptic flow (as we have discussed) and jet quenching (which we have not) suggest that the produced partons re-interact strongly after the CGC stage. The latter was forcefully established 
by the RHIC Deuteron-Gold data. A model which assumes rapid thermalization after the CGC and subsequent ideal 
hydrodynamic evolution with KLN initial conditions was proposed by Hirano and Nara~\cite{HiranoNara}. This apparently 
takes care of the $E_\perp$ and elliptic flow issues. However, several loose ends persist. For instance, even if thermalization 
is rapid, one would expect substantial entropy generation, which is not taken into account in matching the two descriptions. 

In hindsight, despite gaps in our understanding, the evolution in our understanding of heavy ion collisions has exceeded 
even the most optimistic estimates. {\it For instance, much evidence suggests that a strongly interacting quark gluon fluid 
is briefly created at RHIC even if its properties remain to be fully understood~\cite{whitepapers}.} 
The RHIC experimentalists deserve full credit for this by devising ingenious observables 
that have stretched even the most fortunate of models, with most falling like nine pins. Experiments are driving theory, 
disciplining our fancy, and that's how it should be.

\section*{Acknowledgments}

This work has its basis in lectures I delivered at the joint meeting of  HADRON 2004 and  RANP 2004  conferences 
at Angra dos Reis. I wish to thank the conference organizers, in particular Marcelo Chiapparini and Mirian Bracco for 
their painstaking effort. The (short version) of this write up was initiated while I was a guest of the theory group at Saclay.
I want to thank all the members 
of the group for their very kind hospitality. In particular I would like to thank Edmond Iancu and Dionysis Triantafyllopoulos 
for discussions on their recent work and Fran\c cois Gelis for discussions on many topics and for his collaboration on 
some of the topics discussed here. In addition, I would like to thank him for reading the manuscript. 
This revised version has benefited from correspondence with Stephane Munier and careful proof reading by 
Abdel-Nasser Tawfik.  I would also like to thank the members of the theory group at Bielefeld University for their continuing 
support and hospitality during the completion of this work. 
My research is supported in part by DOE Contract No. DE-AC02-98CH10886 and by a research 
grant from the Alexander Von Humboldt Foundation.


\begin{thebibliography}{99}

\bibitem{IV}E. Iancu and R. Venugopalan, hep-ph/0303204, in QGP3, Ed. R. Hwa and X.-N. Wang, 
World Scientific (2004).

\bibitem{RHIC_dA}R.~Debbe  [BRAHMS Collaboration],
J.\ Phys.\ G {\bf 30}, S759 (2004); A.~D.~Frawley  [PHENIX Collaboration],
J.\ Phys.\ G {\bf 30}, S675 (2004); P. Steinberg  [PHOBOS Collaboration],
J.\ Phys.\ G {\bf 30}, S683 (2004); K.~Schweda  [STAR collaboration],
J.\ Phys.\ G {\bf 30}, S693 (2004).

\bibitem{Iancu1}E. Iancu, Proceedings of Hadron 2004/VII RANP, Angra dos Reis, RJ, Brasil, 28 Mar - 3 Apr 2004,  
arXiv:hep-ph/0408228.

\bibitem{McLerran1}L. McLerran, Proceedings of Hadron 2004/VII RANP, Angra dos Reis, RJ, Brasil, 
28 Mar - 3 Apr 2004. 

\bibitem{reviews}A.~H.~Mueller,
arXiv:hep-ph/9911289; L.~McLerran,
Acta Phys.\ Polon.\ B {\bf 34}, 5783 (2003); N.~Armesto,
Acta Phys.\ Polon.\ B {\bf 35}, 213 (2004); 
A.~M.~Stasto,
arXiv:hep-ph/0412084.

\bibitem{GLR}{L.V. Gribov, E.M. Levin, M.G. Ryskin}, Phys. Rept. {\bf 100}, 1 (1983).

\bibitem{MuellerQiu}{A.H. Mueller, J-W. Qiu}, Nucl. Phys. {\bf B} {\bf 268}, 427 (1986).

\bibitem{MV}L.~D.~McLerran and R.~Venugopalan,
Phys.\ Rev.\ D {\bf 49}, 2233 (1994); {\it ibid.}, 3352, (1994); {\it ibid.}, {\bf 50}, 2225 (1994).

\bibitem{JIMWLK}
{J. Jalilian-Marian, A. Kovner, A. Leonidov, H. Weigert}, Nucl. Phys. {\bf B}
  {\bf 504}, 415 (1997); 
{J. Jalilian-Marian, A. Kovner, A. Leonidov, H. Weigert}, Phys. Rev. {\bf D}
  {\bf 59}, 014014 (1999); 
{A. Kovner, G. Milhano, H. Weigert}, Phys. Rev. {\bf D} {\bf 62}, 114005
  (2000); 
{J. Jalilian-Marian, A. Kovner, L.D. McLerran, H. Weigert}, Phys. Rev. {\bf D}
  {\bf 55}, 5414 (1997); 
{E. Iancu, A. Leonidov, L.D. McLerran}, Nucl. Phys. {\bf A} {\bf 692}, 583
  (2001);
{E. Iancu, A. Leonidov, L.D. McLerran}, Phys. Lett. {\bf B} {\bf 510}, 133
  (2001); {E. Ferreiro, E. Iancu, A. Leonidov, L.D. McLerran}, Nucl. Phys. {\bf A} {\bf
  703}, 489 (2002).

\bibitem{GBW1}K. Golec-Biernat, M. W{\"u}sthoff, Phys. Rev. {\bf D} {\bf 59},
  014017 (1999).

\bibitem{IIM}{E. Iancu, K. Itakura, L.D. McLerran}, Nucl. Phys. {\bf A} {\bf 724}, 181
  (2003).

\bibitem{BK}
{I. Balitsky}, Nucl. Phys. {\bf B} {\bf 463}, 99 (1996); 
{Yu.V. Kovchegov}, Phys. Rev. {\bf D} {\bf 61}, 074018 (2000).

\bibitem{Braun}
{M.A. Braun}, Phys. Lett. {\bf B} {\bf 483}, 105 (2000).

\bibitem{ArmestoBraun}
{M.A. Braun}, hep-ph/0101070; 
{N. Armesto, M.A. Braun}, Eur. Phys. J. {\bf C} {\bf 20}, 517 (2001).

\bibitem{GolecStastoMotyka}K.~Golec-Biernat, L.~Motyka and A.~M.~Stasto,
Phys.\ Rev.\ D {\bf 65}, 074037 (2002).

\bibitem{Albacete}J.~L.~Albacete, N.~Armesto, A.~Kovner, C.~A.~Salgado and U.~A.~Wiedemann,
Phys.\ Rev.\ Lett.\  {\bf 92}, 082001 (2004). 

\bibitem{Albacete2}J.~L.~Albacete, N.~Armesto, J.~G.~Milhano, C.~A.~Salgado and U.~A.~Wiedemann,
arXiv:hep-ph/0408216.

\bibitem{RW}K.~Rummukainen and H.~Weigert,
Nucl.\ Phys.\ A {\bf 739}, 183 (2004).

\bibitem{MiklosLarry}M.~Gyulassy and L.~McLerran,
arXiv:nucl-th/0405013.

\bibitem{DL}A.~Donnachie and P.~V.~Landshoff,
Phys.\ Lett.\ B {\bf 296}, 227 (1992).

\bibitem{BFKL}E.~A.~Kuraev, L.~N.~Lipatov and V.~S.~Fadin,
Sov.\ Phys.\ JETP {\bf 45}, 199 (1977)
[Zh.\ Eksp.\ Teor.\ Fiz.\  {\bf 72}, 377 (1977)]; I.~I.~Balitsky and L.~N.~Lipatov,
Sov.\ J.\ Nucl.\ Phys.\  {\bf 28}, 822 (1978)
[Yad.\ Fiz.\  {\bf 28}, 1597 (1978)].

\bibitem{DGLAP}G.~Altarelli and G.~Parisi,
Nucl.\ Phys.\ B {\bf 126}, 298 (1977); V.~N.~Gribov and L.~N.~Lipatov,
Yad.\ Fiz.\  {\bf 15}, 781 (1972)
[Sov.\ J.\ Nucl.\ Phys.\  {\bf 15}, 438 (1972)]; Y.~L.~Dokshitzer,
 ``Calculation Of The Structure Functions For Deep Inelastic Scattering And E+
Sov.\ Phys.\ JETP {\bf 46}, 641 (1977)
[Zh.\ Eksp.\ Teor.\ Fiz.\  {\bf 73}, 1216 (1977)].

\bibitem{Mueller1}A.~H.~Mueller,
Nucl.\ Phys.\ B {\bf 437}, 107 (1995).

\bibitem{Gross}D.~J.~Gross,
Phys.\ Rev.\ Lett.\  {\bf 32}, 1071 (1974).

\bibitem{H1ZEUSQCD}S.~Aid {\it et al.}  [H1 Collaboration],
Phys.\ Lett.\ B {\bf 354}, 494 (1995); A.~M.~Cooper-Sarkar, [ZEUS Collaboration]
arXiv:hep-ph/0110386.

\bibitem{Mueller2}A.~H.~Mueller,
Nucl.\ Phys.\ B {\bf 415}, 373 (1994).

\bibitem{BGP}J.~Bartels, K.~Golec-Biernat and K.~Peters,
Eur.\ Phys.\ J.\ C {\bf 17}, 121 (2000).

\bibitem{NZ}N.~N. Nikolaev and B.~G. Zakharov, Z. \ Phys. \ C{\bf 49}, (1991) 607.

\bibitem{Mueller90}A.~H. Mueller, Nucl.\ Phys. \ B{\bf 335} (1990) 115.

\bibitem{GBW2}{K. Golec-Biernat, M. W{\accent "7F u}sthoff}, Phys. Rev. {\bf D} {\bf 60},
  114023 (1999).
  
\bibitem{MV99}{L.D. McLerran, R. Venugopalan}, Phys. Rev. {\bf D} {\bf 59}, 094002 (1999); R.~Venugopalan,
Acta Phys.\ Polon.\ B {\bf 30}, 3731 (1999).

\bibitem{BGBK}J.~Bartels, K.~Golec-Biernat and H.~Kowalski,
Phys.\ Rev.\ D {\bf 66}, 014001 (2002).

\bibitem{KT}H.~Kowalski and D.~Teaney,
Phys.\ Rev.\ D {\bf 68}, 114005 (2003).

\bibitem{MMS}S.~Munier, A.~M.~Stasto and A.~H.~Mueller,
Nucl.\ Phys.\ B {\bf 603}, 427 (2001).

\bibitem{GRSZ}T.~Rogers, V.~Guzey, M.~Strikman and X.~Zu,
Phys.\ Rev.\ D {\bf 69}, 074011 (2004).

\bibitem{FSLM}L.~Frankfurt, V.~Guzey, M.~McDermott and M.~Strikman,
Phys.\ Rev.\ Lett.\  {\bf 87}, 192301 (2001); E.~Gotsman, E.~Levin, M.~Lublinsky, U.~Maor and E.~Naftali,
arXiv:hep-ph/0302010.

\bibitem{RS}T.~C.~Rogers and M.~I.~Strikman,
arXiv:hep-ph/0410070.

\bibitem{INS}I.~P.~Ivanov, N.~N.~Nikolaev and A.~A.~Savin,
arXiv:hep-ph/0501034.

\bibitem{Polyakov}A. M. Polyakov, {\it Gauge Fields and Strings}, Harwood Publishers, (1987).

\bibitem{GKS}A.~M.~Stasto, K.~Golec-Biernat and J.~Kwiecinski,
Phys.\ Rev.\ Lett.\  {\bf 86}, 596 (2001).

\bibitem{LCreviews}S.~J.~Brodsky, H.~C.~Pauli and S.~S.~Pinsky,
Phys.\ Rept.\  {\bf 301}, 299 (1998); R.~Venugopalan,
arXiv:nucl-th/9808023.

\bibitem{Weinberg}S.~Weinberg, Phys.\ Rev.\ {\bf 150}, 1313, (1966).

\bibitem{Susskind}L.~Susskind, Phys.\ Rev.\ {\bf 165}, 1535, (1968).

\bibitem{Feynman}R.~P.~Feynman, {\it Photon-Hadron Interactions}, Addison Wesley, (1989).

\bibitem{BjKogutSoper}J.~D.~Bjorken, J.~B.~Kogut and D.~E.~Soper,
Phys.\ Rev.\ D {\bf 3}, 1382 (1971).

\bibitem{SR}S.~Jeon and R.~Venugopalan,
Phys.\ Rev.\ D {\bf 70}, 105012 (2004).

\bibitem{JSR}J.~Jalilian-Marian, S.~Jeon and R.~Venugopalan,
Phys.\ Rev.\ D {\bf 63}, 036004 (2001).

\bibitem{JKMW}{J. Jalilian-Marian, A. Kovner, L.D. McLerran, H. Weigert}, Phys. Rev. {\bf D}
  {\bf 55}, 5414 (1997).

\bibitem{Kovchegov}{Yu.V. Kovchegov}, Phys. Rev. {\bf D} {\bf 54}, 5463 (1996); {\it ibid.}, {\bf 55}, 5445 (1997).

\bibitem{AJMV}{A. Ayala, J. Jalilian-Marian, L.D. McLerran, R. Venugopalan}, Phys. Rev. {\bf
  D} {\bf 52}, 2935 (1995); {\it ibid.}, {\bf 53}, 458 (1996).

\bibitem{Weigert}H.~Weigert,
Nucl.\ Phys.\ A {\bf 703}, 823 (2002).

\bibitem{IancuMcLerran}E.~Iancu and L.~D.~McLerran,
Phys.\ Lett.\ B {\bf 510}, 145 (2001).

\bibitem{IIM2}E.~Iancu, K.~Itakura and L.~McLerran,
Nucl.\ Phys.\ A {\bf 724}, 181 (2003).

\bibitem{Mueller3}A.~H.~Mueller,
Nucl.\ Phys.\ B {\bf 643}, 501 (2002).


\bibitem{IIMunier}E.~Iancu, K.~Itakura and S.~Munier,
Phys.\ Lett.\ B {\bf 590}, 199 (2004).

\bibitem{Forshaw}J.~R.~Forshaw, R.~Sandapen and G.~Shaw,
arXiv:hep-ph/0407261.

\bibitem{Forshaw2}J.~R.~Forshaw and G.~Shaw,
JHEP {\bf 0412}, 052 (2004).

\bibitem{CCS}M.~Ciafaloni, D.~Colferai and G.~P.~Salam,
Phys.\ Rev.\ D {\bf 60}, 114036 (1999).

\bibitem{Mueller4}A.~H.~Mueller and D.~N.~Triantafyllopoulos,
Nucl.\ Phys.\ B {\bf 640}, 331 (2002).

\bibitem{Dionysis}D.~N.~Triantafyllopoulos,
Nucl.\ Phys.\ B {\bf 648}, 293 (2003).

\bibitem{Mueller5}{A.H. Mueller}, Nucl. Phys. {\bf A} {\bf 724}, 223 (2003).

\bibitem{LevinTuchin}{E.M. Levin, K. Tuchin}, Nucl. Phys. {\bf B} {\bf 573}, 833 (2000).

\bibitem{IancuMueller}E.~Iancu and A.~H.~Mueller,
Nucl.\ Phys.\ A {\bf 730}, 460 (2004).

\bibitem{MunierPeschanski}S.~Munier and R.~Peschanski,
Phys.\ Rev.\ Lett.\  {\bf 91}, 232001 (2003); Phys.\ Rev.\ D {\bf 69}, 034008 (2004); {\it ibid.}, D {\bf 70}, 077503 (2004).

\bibitem{IMM}E.~Iancu, A.~H.~Mueller and S.~Munier,
arXiv:hep-ph/0410018.

\bibitem{IT}E.~Iancu and D.~N.~Triantafyllopoulos,
arXiv:hep-ph/0411405.

\bibitem{FKPP}R. A. Fischer, Ann.\ Eugenics, {\bf 7}, (1937) 355; A. Kolmogorov, I. Petrovsky, and 
N. Piscounov, {\it Moscow Univ. Bull. Math.}, {\bf A1}, (1937) 1.

\bibitem{MuellerShoshi}A.~H.~Mueller and A.~I.~Shoshi,
Nucl.\ Phys.\ B {\bf 692}, 175 (2004).

\bibitem{MuellerSalam}A~H.~Mueller and G.~P.~Salam,
Nucl.\ Phys.\ B {\bf 475}, 293 (1996).

\bibitem{Salam}G.~P.~Salam,
Nucl.\ Phys.\ B {\bf 449}, 589 (1995); Nucl.\ Phys.\ B {\bf 461}, 512 (1996).

\bibitem{Saarlos}W. van Saarlos, Phys. \ Rep. \ {\bf 386}, (2003) 29; D. Panja, Phys. \ Rep. \ {\bf 393}, (2004) 87.

\bibitem{BrunetDerrida}E.~Brunet and B.~Derrida, Phys. \ Rev. \ {\bf E56}, (1997) 2597; J. \ Stat. \ Phys. {\bf 103}, (2001) 
269.

\bibitem{Baltz}A.~J.~Baltz, F.~Gelis, L.~D.~McLerran and A.~Peshier,
Nucl.\ Phys.\ A {\bf 695}, 395 (2001).

\bibitem{GelisPeshier}F.~Gelis and A.~Peshier,
Nucl.\ Phys.\ A {\bf 697}, 879 (2002).

\bibitem{Bj1}J.~D.~Bjorken,
Lect.\ Notes Phys.\  {\bf 56}, 93 (1976).

\bibitem{Bj2}J.~D.~Bjorken,
Phys.\ Rev.\ D {\bf 27}, 140 (1983).

\bibitem{KMW}A.~Kovner, L.~D.~McLerran and H.~Weigert,
Phys.\ Rev.\ D {\bf 52}, 3809 (1995).

\bibitem{KovRischke}Y.~V.~Kovchegov and D.~H.~Rischke,
Phys.\ Rev.\ C {\bf 56}, 1084 (1997).

\bibitem{GyulassyMcLerran}M.~Gyulassy and L.~D.~McLerran,
Phys.\ Rev.\ C {\bf 56}, 2219 (1997).

\bibitem{GunionBertsch}J.~F.~Gunion and G.~Bertsch,
Phys.\ Rev.\ D {\bf 25}, 746 (1982).

\bibitem{FR}F.~Gelis and R.~Venugopalan,
Phys.\ Rev.\ D {\bf 69}, 014019 (2004).

\bibitem{CiafHaut}S. Catani, M. Ciafaloni, F. Hautmann, Nucl. \ Phys.  \ B {\bf 366}, 135
  (1991).

\bibitem{CollinsEllis}J.~C.~Collins and R.~K.~Ellis,
Nucl.\ Phys.\ B {\bf 360}, 3 (1991).

\bibitem{Shabelski}E.~M.~Levin, M.~G.~Ryskin, Y.~M.~Shabelski and A.~G.~Shuvaev,
Sov.\ J.\ Nucl.\ Phys.\  {\bf 53}, 657 (1991). 

\bibitem{KovMueller}Y.~V.~Kovchegov and A.~H.~Mueller,
Nucl.\ Phys.\ B {\bf 529}, 451 (1998).

\bibitem{KKT} D. Kharzeev, Yu. Kovchegov, K. Tuchin, Phys. \ Rev. \ D {\bf 68}, 094013
  (2003).

\bibitem{BFR1}J.~P.~Blaizot, F.~Gelis and R.~Venugopalan,
Nucl.\ Phys.\ A {\bf 743}, 13 (2004).

\bibitem{BFR2}J.~P.~Blaizot, F.~Gelis and R.~Venugopalan,
Nucl.\ Phys.\ A {\bf 743}, 57 (2004). 

\bibitem{DumitruMclerran}A.~Dumitru and L.~D.~McLerran,
Nucl.\ Phys.\ A {\bf 700}, 492 (2002).


\bibitem{AntreCFSK1}D. Antreasyan, J.W. Cronin, H.J. Frisch, M.J. Shochet, L. Kluberg, P.A.
  Piroue, R.L. Sumner, Phys. \ Rev. \ Lett. {\bf 38}, 112 (1977).

\bibitem{KlubePSAC1}
L. Kluberg, P.A. Piroue, R.L. Sumner, D. Antreasyan, J.W. Cronin, H.J. Frisch,
  M.J. Shochet, Phys. \ Rev. \ Lett. {\bf 38}, 670 (1977).

\bibitem{CroniFSBM1}
J.W. Cronin, H.J. Frisch, M.J. Shochet, J.P. Boymond, R. Mermod, P.A. Piroue,
  R.L. Sumner, Phys. \ Rev. \ D {\bf 11}, 3105 (1975).

\bibitem{KrzywEPS1}
A. Krzywicki, J. Engels, B. Petersson, U. Sukhatme, Phys. \ Lett.  \ B {\bf
  85}, 407 (1979).

\bibitem{QiuS1}J.W. Qiu, G. Sterman, Nucl. \ Phys.  \ B {\bf 353}, 105 (1991).

\bibitem{QiuS2}J.W. Qiu, G. Sterman, Nucl. \ Phys. \ B {\bf 353} 137 (1991).

\bibitem{KopelNST1}B.Z. Kopeliovich, J. Nemchik, A. Schafer, A.V. Tarasov, Phys. \ Rev. \ Lett. {\bf
  88}, 232303 (2002).

\bibitem{Vitev1}I. Vitev, Phys. \ Lett.  \ B {\bf 562}, 36 (2003).

\bibitem{Accar1}A. Accardi, hep-ph/0212148.

\bibitem{AccarG2}A. Accardi, M. Gyulassy, Phys. \ Lett. \ B,  {\bf 586}, 244 (2004).

\bibitem{AccarG3}A. Accardi, M. Gyulassy, nucl-th/0402101.

\bibitem{DumitruJamal}A.~Dumitru and J.~Jalilian-Marian,
Phys.\ Rev.\ Lett.\  {\bf 89}, 022301 (2002).

\bibitem{FrancoisJamal}F.~Gelis and J.~Jalilian-Marian,
Phys.\ Rev.\ D {\bf 66}, 094014 (2002);
Phys.\ Rev.\ D {\bf 66}, 014021 (2002).

\bibitem{FrancoisJamal2}F.~Gelis and J.~Jalilian-Marian,
Phys.\ Rev.\ D {\bf 67}, 074019 (2003).

\bibitem{Jamal}J.~Jalilian-Marian,
Nucl.\ Phys.\ A {\bf 739}, 319 (2004).

\bibitem{BKW}R.~Baier, A.~Kovner and U.~A.~Wiedemann,
Phys.\ Rev.\ D {\bf 68}, 054009 (2003).

\bibitem{JYR}J.~Jalilian-Marian, Y.~Nara and R.~Venugopalan,
Phys.\ Lett.\ B {\bf 577}, 54 (2003).

\bibitem{GSV}V.~Guzey, M.~Strikman and W.~Vogelsang,
Phys.\ Lett.\ B {\bf 603}, 173 (2004).

\bibitem{Debbe1}I.~Arsene {\it et al.}  [BRAHMS Collaboration],
Phys.\ Rev.\ Lett.\  {\bf 93}, 242303 (2004).

\bibitem{PHENIX1}S.~S.~Adler {\it et al.}  [PHENIX Collaboration],
arXiv:nucl-ex/0411054.

\bibitem{STAR1}M.~Ablikim {\it et al.}  [STAR Collaboration],
arXiv:nucl-ex/0408016.

\bibitem{PHOBOS1}B.~B.~Back {\it et al.}  [PHOBOS Collaboration],
arXiv:nucl-ex/0406017.

\bibitem{KLM}D.~Kharzeev, E.~Levin and L.~McLerran,
Phys.\ Lett.\ B {\bf 561}, 93 (2003).

\bibitem{IIT}E.~Iancu, K.~Itakura and D.~N.~Triantafyllopoulos,
Nucl.\ Phys.\ A {\bf 742}, 182 (2004).

\bibitem{STAR2}A.~Ogawa  [STAR Collaboration],
arXiv:nucl-ex/0408004.

\bibitem{KLM2}D.~Kharzeev, E.~Levin and L.~McLerran,
arXiv:hep-ph/0403271.

\bibitem{BMS}R.~Baier, A.~H.~Mueller and D.~Schiff,
Nucl.\ Phys.\ A {\bf 741}, 358 (2004).

\bibitem{Betemps}M.~A.~Betemps and M.~B.~Gay Ducati,
arXiv:hep-ph/0408097.

\bibitem{FriesHwa}R.~C.~Hwa, C.~B.~Yang and R.~J.~Fries,
arXiv:nucl-th/0410111.

\bibitem{Tuchin}K.~Tuchin,
Phys.\ Lett.\ B {\bf 593}, 66 (2004).

\bibitem{KopRauf}B. Z. Kopeliovich, J. Raufeisen and A. V. Tarasov, Phys.\ Rev. \ C {\bf 62} 035204 (2000).

\bibitem{Raufeisen}J.~Raufeisen, hep-ph/0409324.

\bibitem{ShafNik}N. N. Nikolaev, G. Piller and B. G. Zakharov, J. \ Exp. \ Theor. \ Phys. {\bf 81}, 851 (1995); Z. \ Phys. \ 
A {\bf 354}, 99 (1996); N.~N.~Nikolaev and W.~Schafer,
arXiv:hep-ph/0411365.

\bibitem{GoncalvesMachado}V.~P.~Goncalves and M.~V.~T.~Machado,
arXiv:hep-ph/0410012.

\bibitem{FFR}H. Fujii, F. Gelis and R. Venugopalan, in preparation. 

\bibitem{Kopeliovich}B.~Z.~Kopeliovich, J.~Raufeisen, A.~V.~Tarasov and M.~B.~Johnson,
Phys.\ Rev.\ C {\bf 67}, 014903 (2003).

\bibitem{KovWied}A.~Kovner and U.~A.~Wiedemann,
Phys.\ Rev.\ D {\bf 64}, 114002 (2001).

\bibitem{Balitsky2}I.~Balitsky,
arXiv:hep-ph/0409314.

\bibitem{AR}A.~Krasnitz and R.~Venugopalan,
Nucl.\ Phys.\ B {\bf 557}, 237 (1999); Phys.\ Rev.\ Lett.\  {\bf 84}, 4309 (2000); Phys.\ Rev.\ Lett.\  {\bf 86}, 1717 (2001).

\bibitem{AYR}A.~Krasnitz, Y.~Nara and R.~Venugopalan,
Phys.\ Rev.\ Lett.\  {\bf 87}, 192302 (2001); Nucl.\ Phys.\ A {\bf 717}, 268 (2003).

\bibitem{Lappi}T.~Lappi, Phys.\ Rev.\ C {\bf 67}, 054903 (2003).

\bibitem{AYR2}A.~Krasnitz, Y.~Nara and R.~Venugopalan,
Nucl.\ Phys.\ A {\bf 727}, 427 (2003).

\bibitem{AYR3}A.~Krasnitz, Y.~Nara and R.~Venugopalan,
Phys.\ Lett.\ B {\bf 554}, 21 (2003).

\bibitem{Mueller6}A.~H.~Mueller,
Phys.\ Lett.\ B {\bf 475}, 220 (2000).

\bibitem{BV}J.~Bjoraker and R.~Venugopalan,
Phys.\ Rev.\ C {\bf 63}, 024609 (2001).

\bibitem{SerreauSchiff}J.~Serreau and D.~Schiff,
JHEP {\bf 0111}, 039 (2001).

\bibitem{BMSS}R.~Baier, A.~H.~Mueller, D.~Schiff and D.~T.~Son,
Phys.\ Lett.\ B {\bf 502}, 51 (2001).

\bibitem{BMSS2}R.~Baier, A.~H.~Mueller, D.~Schiff and D.~T.~Son,
Phys.\ Lett.\ B {\bf 539}, 46 (2002).

\bibitem{Stan}S.~Mrowczynski,
Phys.\ Lett.\ B {\bf 314}, 118 (1993).

\bibitem{AML}P.~Arnold, J.~Lenaghan and
 G.~D.~Moore, JHEP {\bf 0308}, 002 (2003).

\bibitem{RS1}P.~Romatschke and M.~Strickland,
Phys.\ Rev.\ D {\bf 68}, 036004 (2003).

\bibitem{Boyanovsky}D.~Boyanovsky, C.~Destri and H.~J.~de Vega,
Phys.\ Rev.\ D {\bf 69}, 045003 (2004); D.~Boyanovsky, H.~J.~de Vega, R.~Holman, S.~P.~Kumar, R.~D.~Pisarski and J.~Salgado,
arXiv:hep-ph/9810209.

\bibitem{Dietrich}D.~D.~Dietrich,
Phys.\ Rev.\ D {\bf 70}, 105009 (2004); {\it ibid}., {\bf 68}, 105005 (2003).

\bibitem{FTK}F.~Gelis, K.~Kajantie and T.~Lappi,
arXiv:hep-ph/0409058.

\bibitem{KLN}D.~Kharzeev and M.~Nardi,
Phys.\ Lett.\ B {\bf 507}, 121 (2001); D.~Kharzeev and E.~Levin,
Phys.\ Lett.\ B {\bf 523}, 79 (2001); D.~Kharzeev, E.~Levin and M.~Nardi,
arXiv:hep-ph/0111315.

\bibitem{HiranoNara}T.~Hirano and Y.~Nara,
Nucl.\ Phys.\ A {\bf 743}, 305 (2004).

\bibitem{whitepapers}I.~Arsene {\it et al.}  [BRAHMS Collaboration],
arXiv:nucl-ex/0410020; K.~Adcox {\it et al.}  [PHENIX Collaboration],
arXiv:nucl-ex/0410003, and presentations by W. Busza  [PHOBOS] and 
R. Snellings [STAR] at RBRC workshop {\it New Discoveries at RHIC}, BNL, May 14th-15th, 2004.

\end{thebibliography}
\end{document}